\documentclass[10pt]{article}
\usepackage{amsfonts}
\usepackage{amsmath}
\usepackage{amssymb}
\usepackage{graphicx}
\usepackage{subfigure}

\usepackage{bbold}

\usepackage[latin1]{inputenc}% Caracteres acentuados

\newcommand{\topico}[1]{\subsubsection*{#1}} 
\newcommand{\N}{{\mathbb{N}}}
\newcommand{\R}{{\mathbb{R}}}
\newcommand{\C}{{\mathbb{C}}}

\newcommand{\UM}{\mathbb{1}}
\newcommand{\ZERO}{\mathbb{0}}

\newcommand{\Mat}{\mathrm{Mat}\,}

\newcommand{\meucoloneqq}{\mathrel{\mathop:}=}
\newcommand{\meueqqcolon}{=\mathrel{\mathop:}}
\newcommand{\defi}{\meucoloneqq}
\newcommand{\ifed}{\meueqqcolon}
\newcommand{\lan}{{\langle }}
\newcommand{\biglan}{{\big\langle }}
\newcommand{\Biglan}{{\Big\langle }}

\newcommand{\ran}{{\rangle }}
\newcommand{\bigran}{{\big\rangle }}
\newcommand{\Bigran}{{\Big\rangle }}

\newcommand{\Lcolchete}{{{\Big[\!\! \Big[}}}
\newcommand{\Rcolchete}{{{\Big]\!\! \Big]}}}
\newcommand{\diag}{\mathrm{diag}\,}

\newcommand{\calA}{{\mathcal A}}
\newcommand{\calH}{{\mathcal H}}
\newcommand{\calM}{{\mathcal M}}
\newcommand{\calS}{{\mathcal S}}
\newcommand{\calV}{{\mathcal V}}
\newtheorem{theorem}{Theorem}[section]
\newtheorem{lemma}[theorem]{Lemma}
\newtheorem{proposition}[theorem]{Proposition}

\newcommand{\eps}{\epsilon}

\newcommand{\Fullbox}{\bigskip\hfill{\rule{2.5mm}{2.5mm}}}
\newcommand{\EndofStatement}{\samepage\smallskip\hfill\Box}

\newcommand{\QED}{\Fullbox}
\newcommand{\Proof}[1]{{

\medskip
\noindent{\bf\sf Proof{#1}.}}} 

\newcommand{\Ran}{\mathrm{Ran}\,}
\newcommand{\Ker}{\mathrm{Ker}\,}

\begin{document}

\begin{center}

\begin{Large}
{\bf 
The Moore-Penrose Pseudoinverse. A Tutorial Review of the Theory
}
\end{Large}

J.\ C.\ A.\ Barata\footnote{e-mail: jbarata@if.usp.br} 
and 
M.\ S.\ Hussein\footnote{e-mail: hussein@if.usp.br}

Instituto de F\'{i}sica, Universidade de S\~{a}o Paulo,
  C.P.\ 66318, 05314-970 S\~{a}o Paulo, SP, Brazil

\end{center}

\abstract{ In the last decades the Moore-Penrose pseudoinverse has
  found a wide range of applications in many areas of Science and
  became a useful tool for physicists dealing, for instance, with
  optimization problems, with data analysis, with the solution of
  linear integral equations, etc.  The existence of such applications
  alone should attract the interest of students and researchers in the
  Moore-Penrose pseudoinverse and in related subjects, like the
  singular values decomposition theorem for matrices.
  In this note we present a tutorial review of the theory of the
  Moore-Penrose pseudoinverse. We present the first definitions and
  some motivations and, after obtaining some basic results, we center
  our discussion on the Spectral Theorem and present an
  algorithmically simple expression for the computation of the
  Moore-Penrose pseudoinverse of a given matrix. We do not claim
  originality of the results. We rather intend to present a complete
  and self-contained tutorial review, useful for those more devoted to
  applications, for those more theoretically oriented and for those
  who already have some working knowledge of the subject.}

\section{Introduction, Motivation and Notation}

In this paper we present a self-contained review of some of the basic
results on the so-called Moore-Penrose pseudoinverse of matrices, a
concept that generalizes the usual notion of inverse of a square
matrix, but that is also applicable to singular square matrices or
even to non-square matrices. This notion is particularly useful in
dealing with certain linear least squares problems, as we shall
discuss in Section
\ref{sec:A-Pseudo-Inversa-de-Moore-Penrose-e-Problemas-de-Optimizacao-Linear},
i.e., problems where one searches for an optimal approximation for
solutions of linear equations like $Ax=y$, where $A$ is a given
$m\times n$ matrix, $y$ is a given column vector with $m$ components
and the unknown $x$, a column vector with $n$ components, is the
searched solution. In many situations, a solution is non-existing or
non-unique, but one asks for a vector $x$ such that the norm of the
difference $Ax-y$ is the smallest possible (in terms of least
squares).

Let us be a little more specific.
Let $A\in\Mat(\C, \; m, \; n)$ (the set of all complex $m\times n$
matrices) and $y\in \C^m$ be given and consider the problem of finding
$x\in \C^n$ satisfying the linear equation
\begin{equation}
  Ax \; = \; y
\; .
\label{eq:LuybiUYtvtRytfr-00}
\end{equation}
If $m=n$ and $A$ has an inverse, the (unique) solution is, evidently,
$x=A^{-1}y$. In other cases the solution may not exist or may not
be unique. We can, however, consider the alternative problem of
finding the set of all vectors $x'\in \C^n$ such that the Euclidean
norm $\|Ax'-y\|$ reaches its least possible value. This set is called
the {\em minimizing set} of the linear problem
(\ref{eq:LuybiUYtvtRytfr-00}). Such vectors $x'\in \C^n$ would be the
best approximants for the solution of (\ref{eq:LuybiUYtvtRytfr-00}) in
terms of the Euclidean norm, i.e., in terms of ``least squares''.  As
we will show in Theorem \ref{teor:OptimizacaoLinear}, the
Moore-Penrose pseudoinverse provides this set of vectors $x'$ that
minimize $\|Ax'-y\|$: it is the set
\begin{equation}
 \Big\{  A^+y +\big( \UM_n -A^+A\big) z, \; z\in\C^n \Big\} 
\; ,
\label{eq:conjuntominimizante-deoptimizacaolinear-00}
\end{equation}
where $A^+\in \Mat(\C, \; n, \; m)$ denotes the Moore-Penrose
pseudoinverse of $A$.  An important question for applications is to
find a general and algorithmically simple way to compute $A^+$. The
most common approach uses the singular values decomposition and is
described in Appendix
\ref{app:Existencia-e-Decomposicao-em-Valores-Singulares}.  Using the
Spectral Theorem and Tikhonov's regularization method we show that
$A^+$ can be computed by the algorithmically simpler formula
\begin{equation}
A^+ 
\; = \;
\sum_{{b=1}\atop{\beta_b\neq 0}}^{s}
\;\;
\frac{1}{\beta_b}
\left(
\prod_{{l=1}\atop {l\neq b}}^s
  \big(\beta_b-\beta_l\big)^{-1}
\right)
\;\;
\left[
\prod_{{l=1}\atop {l\neq b}}^s 
\Big( A^*A-\beta_l\UM_n \Big)
\right]
A^*
\; ,
\label{eq:reppreudoespectraldeAMAIS-2-00}
\end{equation}
where $A^*$ denotes the adjoint matrix of $A$ and $\beta_k$, $k=1,\;
\ldots , \; s$, are the distinct eigenvalues of $A^*A$ (the so-called
{\em singular values} of $A$). See Theorem
\ref{teor:representacaoquaseespectaldeAMAIS} for a more detailed
statement. One of the aims of this paper is to present a proof of
(\ref{eq:reppreudoespectraldeAMAIS-2-00}) by combining the spectral
theorem with the a regularization procedure due to Tikhonov
\cite{Tikhonov2,TikhonovArsenin}.

\topico{Some applications of the Moore-Penrose pseudoinverse}

Problems involving the determination of the minimizing set of
(\ref{eq:LuybiUYtvtRytfr-00}) are always present when the number of
unknowns exceeds the number of values provided by measurements. Such
situations occur in many areas of Applied Mathematics, Physics and
Engineering, ranging from imaging methods, like 
MRI (magnetic resonance imaging)
\cite{SodicksonMcKenzie,VanDeWalle-et-alli,HabibAmmari},
fMRI (functional MRI) \cite{XingfengLi-et-alii,Lohmann-et-alii},
PET (positron emission tomography) \cite{Stefanescu-et-allii,BerteroBoccacci} and
MSI (magnetic source imaging)
\cite{Jia-Zhu-Wang-et-allii-I,GencerWilliamson,Jia-Zhu-Wang-et-allii-II},
to seismic inversion problems \cite{PageCustodioArchuletaCarlson,AtzoriAntonioli}.

The Moore-Penrose pseudoinverse and/or the singular values
decomposition (SVD) of matrices (discussed in Appendix
\ref{app:Existencia-e-Decomposicao-em-Valores-Singulares}) are also
employed in data analysis, as in the treatment of
electroencephalographic source localization \cite{Pascual-Marqui} and
in the so-called Principal Component Analysis (PCA). Applications of
this last method to astronomical data analysis can be found in
\cite{BediniHerranzSalernoBaccigalupiKuruoguzTonazzini,HennigChristlieb,RicciSteinerMenezes-NGC7097,SteinerMenezesRicciOliveira-PCATomography}
and applications to gene expression analysis can be found in
\cite{WallRechtsteinerRocha,TroyanskayaCantorSherlockBrownHastieTibshiraniBotsteinAltman}.
Image compression algorithms using SVD are known at least since
\cite{AndrewsPatterson} and digital image restoration using the
Moore-Penrose pseudoinverse have been studied in
\cite{ChountasisKatsikisPappas-I,ChountasisKatsikisPappas-II}.

% ******************

%
Problems involving the determination of the minimizing set of
(\ref{eq:LuybiUYtvtRytfr-00}) also occur, for instance, in certain
numerical algorithms for finding solutions of linear Fredholm integral
equations of the first kind:
$$
\int_a^b k(x, \; y)\, u(y)\;dy 
\; = \; 
f(x)
\;,
%\label{eq:Fredholm-of-the-first-kind-0}
$$
where $-\infty<a<b<\infty$ and where $k$ and $f$ are given functions.
See Section \ref{sec:RegularizacaodeTikhonov} for a further discussion
of this issue. For an introductory account on integral equations, rich
in examples and historical remarks, see
\cite{Groetsch-IntegralEquations}.

Even this short list of applications should convince a student of
Physics or Applied Mathematics of the relevance of the Moore-Penrose
pseudoinverse and related subjects and our main objective is to
provide a self-contained introduction to the required theory.

\topico{Organization}

In Section \ref{sec:MPPI-Fist-Properties}
we present the definition of the Moore-Penrose pseudoinverse and
obtain its basic properties.
In Section \ref{sec:OutrasPropriedadesdaPseudo-InversadeMoore-Penrose}
we further develop the theory of the  Moore-Penrose pseudoinverses.
In Section \ref{sec:RegularizacaodeTikhonov}
we describe 
Tikhonov's regularization method for the computation of 
Moore-Penrose pseudoinverses and present a first proof of existence.
Section
\ref{sec:A-Pseudo-Inversa-de-Moore-Penrose-e-o-Teorema-Espectral}
collects the previous results and derives expression
(\ref{eq:reppreudoespectraldeAMAIS-2-00}), based on the Spectral
Theorem, for the computation of Moore-Penrose pseudoinverses. This
expression is algorithmically simpler than the usual method based on
the singular values decomposition (described in Appendix
\ref{app:Existencia-e-Decomposicao-em-Valores-Singulares}).
In Section
\ref{sec:A-Pseudo-Inversa-de-Moore-Penrose-e-Problemas-de-Optimizacao-Linear}
we show the relevance of the Moore-Penrose pseudoinverse for the
solution of linear least squares problems, its main motivation.
In Appendix
\ref{app:A-Brief-Review-of-Hilbert-Space-Theory-and-Linear-Algebra} we
present a self-contained review of the results on Linear Algebra and
Hilbert space theory, not all of them elementary, that we need in the
main part of this paper.
In Appendix \ref{app:Existencia-e-Decomposicao-em-Valores-Singulares}
we approach the existence problem of the Moore-Penrose pseudoinverse
by using the usual singular values decomposition method.

\topico{Notation and preliminary definitions}

In the following we fix the notation utilized throughout the paper.
We denote $\C^n$ the vector space of all $n$-tuples of complex
numbers: $\C^n\defi\left\{ \left(\begin{smallmatrix}z_1\\
      \vdots\\z_n\end{smallmatrix}\right), \; \mbox{ with } z_k\in\C
  \mbox{ for all } k=1, \; \ldots, \; n\right\}$. We denote the usual
scalar product in $\C^n$ by $\lan\cdot, \; \cdot\ran_\C$ or simply by 
 $\lan\cdot, \; \cdot\ran$, where
for $z= \left(\begin{smallmatrix}z_1\\
      \vdots\\z_n\end{smallmatrix}\right)\in\C^n$ and
$w= \left(\begin{smallmatrix}w_1\\
      \vdots\\w_n\end{smallmatrix}\right)\in\C^n$, we have
$$
\lan z, \; w \ran_\C
\; \equiv \;
\lan z, \; w \ran
\; \defi \;
\sum_{k=1}^n\overline{z_k}w_k
\;.
$$
Note that this scalar product is linear in the second argument and
anti-linear in the first, in accordance with the convention adopted in
Physics. Two vectors $u$ and $v\in \C^n$ are said to be orthogonal
according to the scalar product $\lan \cdot , \; \cdot \ran$ if $\lan
u , \; v \ran=0$. If $W\subset \C^n$ is a subspace of $\C^n$ we denote
by $W^\perp$ the subspace of $\C^n$ composed by all vectors orthogonal
to all vectors of $W$. The usual norm of a vector $z\in\C^n$ 
will be denoted by $\|z\|_\C$ or simply by $\|z\|$ and is defined by
$\|z\|_\C\equiv\|z\|=\sqrt{\lan z, \; x\ran}$. It is well known that 
$\C^n$ is a Hilbert space with respect to the usual scalar product.

The set of all complex $ m\times n$ matrices ($m$ rows and $n$
columns) will be denoted by $\Mat (\C , \; m, \; n)$. The set of all
square $ n\times n$ matrices with complex entries will be denoted by
$\Mat (\C , \; n)$.

The identity matrix will be denoted by  $\UM$. Given 
$A\in \Mat (\C , \; m, \;n)$ we denote by $A^T$ 
element  of  $\Mat (\C , \; n, \;m)$ whose matrix elements are
$(A^T)_{ij} = A_{ji}$ for all $i\in\{1, \; \ldots , \; n\}$,
$j\in\{1, \; \ldots , \; m\}$. The matrix $A^T$ is said to be the
transpose of $A$. It is evident that $(A^T)^T=A$ and that 
$(AB)^T=B^TA^T$ for all $A\in \Mat (\C , \; m, \;n)$ and
$B\in \Mat (\C , \; n, \;p)$.

If $A\in\Mat(\C, \; m, \; n)$, then its adjoint $A^*\in\Mat(\C, \; n,
\; m)$ is defined as the matrix whose matrix elements $(A^*)_{ij}$ are
given by $\overline{A_{ji}}$ for all $0\leq i \leq n$ and $0\leq j
\leq m$.

Given a set $\alpha_1, \; \ldots , \; \alpha_n $ of complex numbers we
denote by $\diag (\alpha_1, \; \ldots , \; \alpha_n )\in \Mat (\C, \;
n)$ the diagonal matrix  whose $k$-th diagonal
entry is $\alpha_k$:
$$
\big(\diag (\alpha_1, \; \ldots , \; \alpha_n )\big)_{ij} \; = \; 
\left\{
\begin{array}{ll}
 \alpha_i, & \mbox{for } i=j \; , \\
  0, & \mbox{for } i\neq j \; .
\end{array}
\right.  
$$

The spectrum of a square matrix $A\in \Mat(\C, \; n)$ coincides with
the set of its eigenvalues (see the definitions in Appendix
\ref{app:A-Brief-Review-of-Hilbert-Space-Theory-and-Linear-Algebra})
and will be denoted by $\sigma(A)$.

We denote by $\ZERO_{a, \; b}\in \Mat (\C , \; a, \;b)$ the $a\times
b$ whose matrix elements are all zero. We denote by $\UM_{l}\in \Mat
(\C , \; l)$ the $l\times l$ identity matrix. If no danger of
confusion is present, we will simplify the notation and write $\ZERO$
and $\UM$ instead of $\ZERO_{a, \; b}$ and $\UM_{l}$, respectively.
We will also employ the following definitions: for $m, \; n\in\N$, let
$I_{m, \; m+n}\in \Mat(\C, \; m, \; m+n)$ and $J_{m+n, \; n}\in \Mat(\C,
\; m+n, \; n)$ be given by
\begin{equation}
I_{m, \; m+n} 
\; \defi \; 
\begin{pmatrix}
\UM_m & \ZERO_{m, \; n}
\end{pmatrix}
\qquad \mbox{ and } \qquad
J_{m+n, \; n} 
\; \defi \; 
\begin{pmatrix}
\UM_n \\ \ZERO_{m, \; n}
\end{pmatrix}
\; .
\label{eq:MatrizesIJeAlinha-1}
\end{equation}
The corresponding transpose matrices are
\begin{equation}
\big(I_{m, \; m+n}\big)^T 
\; \defi \; 
\begin{pmatrix}
\UM_m \\ \ZERO_{n, \; m}
\end{pmatrix}
\; = \; J_{m+n, \; m} 
\qquad \mbox{ and } \qquad
\big(J_{ m+n , \; n}\big)^T 
\; \defi \; 
\begin{pmatrix}
\UM_n & \ZERO_{n, \; m}
\end{pmatrix}
\; = \; I_{n, \; m+n}
\; .
\label{eq:MatrizesIJeAlinha-2}
\end{equation}
The following useful identities will be used bellow:
\begin{eqnarray}
I_{m, \; m+n} \, \big(I_{m, \; m+n}\big)^T   & = & 
I_{m, \; m+n} J_{m+n, \; m} \; = \; 
\UM_m \;,
\label{eq:MatrizesIJeAlinha-3}
\\
\big(J_{ m+n , \; n}\big)^T J_{ m+n , \; n} & = & 
I_{n, \;  m+n } J_{ m+n , \; n} \; = \; 
\UM_n \;,
\label{eq:MatrizesIJeAlinha-4}
\end{eqnarray}

For each $A\in \Mat(\C, \; m, \; n)$ we can associate a 
square matrix $A'\in \Mat(\C, \; m+n)$ given by
\begin{equation}
A' 
\; \defi \; 
\big(I_{m, \; m+n}\big)^T A \big(J_{ m+n , \; n}\big)^T
\; = \; 
J_{m+n, \; m} A I_{ n, \; m+n}
\; = \; 
\begin{pmatrix}
A & \ZERO_{m, \; m}
\\
\ZERO_{n, \; n} & \ZERO_{n, \; m}
\end{pmatrix}
\;.
\label{eq:MatrizesIJeAlinha-5}
\end{equation}
As one easily checks, we get from
(\ref{eq:MatrizesIJeAlinha-3})--(\ref{eq:MatrizesIJeAlinha-4})
the useful relation
\begin{equation}
A 
\; = \; 
I_{m, \; m+n} A' J_{ m+n , \; n}
\; .
\label{eq:MatrizesIJeAlinha-6}
\end{equation}

The canonical basis of vectors in $\C^n$ is
\begin{equation}
{\mathbf e}_1 \; = \; 
\begin{pmatrix}
1 \\ 0 \\ 0\\ \vdots \\ 0  
\end{pmatrix}  , \qquad
{\mathbf e}_2 \; = \; 
\begin{pmatrix}
0 \\ 1 \\ 0 \\ \vdots \\ 0  
\end{pmatrix}  , \qquad
\ldots , \qquad
{\mathbf e}_n \; = \; 
\begin{pmatrix}
0 \\0\\  \vdots \\ 0 \\1  
\end{pmatrix}  \; ,
\label{eq:ABaseCanonica-1}
\end{equation}
Let $ x^1, \; \ldots , \; x^n$ be vectors, represented in the
canonical basis as
$$
x^a \; = \; 
\begin{pmatrix}
x^a_1 \\ \vdots \\ x^a_n  
\end{pmatrix}  \;  .
$$
We will denote by
$
\Lcolchete x^1, \; \ldots , \; x^n  \Rcolchete
$
the $ n\times n$ constructed in such a way that its $a$-th column is
the vector  $x^a$, that means,
\begin{equation}
\Lcolchete x^1, \; \ldots , \; x^n  \Rcolchete 
\; = \; 
\begin{pmatrix}
x^1_1 & \cdots & x^n_1 \\
 \vdots & \ddots & \vdots \\
x^1_n & \cdots & x^n_n 
\end{pmatrix} \;  .
\label{eq:notacaoparamatrizgeradaporcolunas}
\end{equation}
It is obvious that
$
\UM  = \Lcolchete {\mathbf e}_1 , \; \ldots , \; {\mathbf e}_n \Rcolchete 
$.
With this notation we write
\begin{equation}
\label{BgerV}
B \Lcolchete x^1, \; \ldots , \; x^n  \Rcolchete 
\; = \; 
\Lcolchete Bx^1, \; \ldots , \; Bx^n  \Rcolchete  \; ,
\end{equation}
for any $B\in \Mat(\C, \; m, \; n)$, as one easily checks.
Moreover, if $D$ is a diagonal matrix
$D=\diag  (d_1, \; \ldots , \; d_n )$, then
\begin{equation}
\label{eq:gerVmalD}
\Lcolchete x^1, \; \ldots , \; x^n  \Rcolchete \, D
\; = \; 
\Lcolchete d_1 x^1, \; \ldots , \; d_n x^n  \Rcolchete\; .
\end{equation}

If $v_1 , \; \ldots , \; v_k$ are elements of a complex vector space
$V$, we denote by $ [v_1 , \; \ldots , \; v_k] $ the subspace
generated $v_1 , \; \ldots , \; v_k$, i.e., the collection of all
linear combinations of the $v_1 , \; \ldots , \; v_k$: $ [v_1 , \;
\ldots , \; v_k] \defi \Big\{ \alpha_1 v_1 + \cdots + \alpha_k v_k, \;
\; \alpha_1, \; \ldots , \; \alpha_k \in \C \Big\} $.

More definitions and general results can be found in Appendix
\ref{app:A-Brief-Review-of-Hilbert-Space-Theory-and-Linear-Algebra}.

\section{The Moore-Penrose Pseudoinverse. 
Definition and First Properties}
\label{sec:MPPI-Fist-Properties}

In this section we define the notion of a Moore-Penrose pseudoinverse
and study its uniqueness. The question of the existence of the
Moore-Penrose pseudoinverse of a given matrix is analyzed in other
sections.

\topico{Generalized inverses, or pseudoinverses}

Let $m, \; n\in\N$ and let $A\in\Mat(\C, \; m \; ,n)$ be a $m\times n$
matrix (not necessarily a square matrix). A matrix $B\in\Mat(\C, \; n,
\; m)$ is said to be a {\em generalized inverse}, or a {\em
  pseudoinverse}, of $A$ if it satisfies the following conditions:
\begin{enumerate}
\item $ABA=A$,
\item $BAB=B$.
\end{enumerate}
If $A\in \Mat(\C, \; n)$ is a non-singular square matrix, its inverse
$A^{-1}$ satisfies trivially the defining properties of the
generalized inverse above.  We will prove later that every matrix
$A\in \Mat(\C, \; m \; ,n)$ has at least one generalized inverse, namely,
the Moore-Penrose pseudoinverse. The general definition above is not
enough to guarantee uniqueness of the generalized inverse of any
matrix $A\in\Mat(\C, \; m \; ,n)$. 

The definition above is too wide to be useful and it is convenient to
narrow it in order to deal with certain specific problems. In what
follows we will discuss the specific case of the Moore-Penrose
pseudoinverse and its application to optimization of linear least
squares problems.

\topico{Defining the Moore-Penrose pseudoinverse}

Let $m, \; n\in\N$ and let 
$A\in\Mat(\C, \; m \; ,n)$. A matrix $A^+\in\Mat(\C, \; n, \; m)$ is
said to be a {\em  Moore-Penrose pseudoinverse} of $A$ if it
satisfies the following conditions:
\begin{enumerate}
\item 
$AA^+A=A$,
\item 
$A^+AA^+=A^+$,
\item 
$AA^+ \in \Mat(\C, \; m)$ and $A^+A \in \Mat(\C, \; n)$ are self-adjoint.
\end{enumerate}

It is easy to see again that if $A\in \Mat(\C, \; n)$ is non-singular,
then its inverse satisfies all defining properties of a Moore-Penrose
pseudoinverse. 

The notion of Moore-Penrose pseudoinverse was introduced by E.\ H.\
Moore \cite{Moore1} in 1920 and
rediscovered by R.\ Penrose \cite{Penrose1,Penrose2} in
1955.
The Moore-Penrose pseudoinverse is a useful concept in dealing with
optimization problems, as the determination of a ``least squares''
solution of linear systems.  We will treat such problems later (see
Theorem \ref{teor:OptimizacaoLinear}), after dealing with the question
of uniqueness and existence of the Moore-Penrose pseudoinverse.

\topico{The uniqueness of the Moore-Penrose pseudoinverse}

We will first show the uniqueness of the Moore-Penrose pseudoinverse
of a given matrix $A\in\Mat(\C, \; m, \; n)$, assuming its existence.

Let $A^+\in\Mat(\C, \; n,\;m)$ be a Moore-Penrose pseudoinverse
$A\in\Mat(\C, \; m,\;n)$ and let $B\in\Mat(\C, \; n,\;m)$ be another
Moore-Penrose pseudoinverse of $A$, i.e., such that $ABA=A$, $BAB=B$
with $AB$ and $BA$ self-adjoint.  Let $M_1\defi
AB-AA^+=A\big(B-A^+\big)\in \Mat(\C, \; m)$. By the hypothesis, $M_1$
is self-adjoint (since it is the difference of two self-adjoint
matrices) and $(M_1)^2=\big(AB-AA^+\big)A\big(B-A^+\big)=
\big(ABA-AA^+A\big)\big(B-A^+\big)=(A-A)\big(B-A^+\big) =0$. Since
$M_1$ is self-adjoint, the fact that $(M_1)^2=0$ implies that $M_1=0$,
since for all $x\in\C^m$ one has $\|M_1x\|^2=\lan M_1x, \,
M_1x\ran=\biglan x, \, (M_1)^2x\bigran=0$, implying $M_1=0$.  This
showed that $AB=AA^+$. Following the same steps we can prove that
$BA=A^+A$ (consider the self-adjoint matrix $M_2\defi BA-A^+A\in
\Mat(\C, \; n)$ and proceed as above).  Now, all this implies that $A^+
= A^+AA^+=A^+\big(AA^+\big)=A^+AB=\big(A^+A\big)B=BAB=B$, thus establishing
uniqueness.

As we already commented, if $A\in \Mat(\C, \; n)$ is a non-singular
square matrix, its inverse $A^{-1}$ trivially satisfies the defining
conditions of the Moore-Penrose pseudoinverse and, therefore, 
we have in this case $A^+=A^{-1}$ as the unique Moore-Penrose
pseudoinverse of $A$. 
It is also evident from the definition that for $\ZERO_{mn}$, the
$m\times n$ identically zero matrix, one has
$(\ZERO_{mn})^+=\ZERO_{nm}$.

\topico{Existence of the Moore-Penrose pseudoinverse}

We will present two proofs of the existence of the Moore-Penrose
pseudoinverse $A^+$ for an arbitrary matrix $A\in \Mat(\C, \; m, \;
n)$.  Both proofs produce algorithms for the explicit computation of
$A^+$. The first one will be presented in Section
\ref{sec:RegularizacaodeTikhonov} (Theorems
\ref{teor:regularizacaodeTychonovparaAMAIS} and
\ref{teor:representacaoquaseespectaldeAMAIS}) and will follow from
results presented below. Expressions
(\ref{eq:reppreudoespectraldeAMAIS-1}) and
(\ref{eq:reppreudoespectraldeAMAIS-2}) furnish explicit expressions
for the computation of $A^+$ in terms of $A$, $A^*$ and the eigenvalues
of $AA^*$ or $A^* A$ (i.e., the singular values of $A$).

The second existence proof will be presented in Appendix
\ref{app:Existencia-e-Decomposicao-em-Valores-Singulares} and relies
on the singular values decomposition presented in Theorem
\ref{teo:teoremadadecomposicaoemvaloressingularesdematrizes}. For
this proof, the preliminary results presented below are not
required. This second proof is the one more frequently found in the
literature, but we believe that expressions
(\ref{eq:reppreudoespectraldeAMAIS-1}) and
(\ref{eq:reppreudoespectraldeAMAIS-2}) provide an algorithmically
simpler way for the determination of the Moore-Penrose pseudoinverse
of a given matrix.

\topico{Computing the Moore-Penrose pseudoinverse in some special cases}

If $A\in\Mat(\C, \; m, \; 1)$,
$A=\left(\begin{smallmatrix}a_1 \\ \vdots \\
    a_m\end{smallmatrix}\right)$, a non-zero column vector, then one
can easily verify that $A^+=\frac{1}{\|A\|^2}A^* =
\frac{1}{\|A\|^2}\left(\begin{smallmatrix}\overline{a_1}\;, &\ldots
    , & \overline{a_m}\end{smallmatrix}\right)$, where
$\|A\|=\sqrt{|a_1|^2+ \cdots +|a_m|^2}$.
In particular, if $z\in\C$, then $(z)^+=\left\{\begin{array}{ll}0, & z=0\\
    \frac{1}{z},&z\neq 0\end{array}\right.$, by taking $z$ as an
element of $\Mat(\C, \; 1, \; 1)$.

This can be further generalized. If $A\in\Mat(\C, \; m, \; n)$ and
$(AA^*)^{-1}$ exists, then 
\begin{equation}
A^+ \; = \; A^*\big(AA^*\big)^{-1}
\; ,
\label{eq:pseudoinversa-AAeEXTinversivel-1}
\end{equation}
because we can
readly verify that the r.h.s.\ satisfies the defining conditions of
$A^+$. Analogously, if $(A^*A)^{-1}$ exists, one has 
\begin{equation}
A^+ \; = \; \big(A^*A\big)^{-1}A^*
\; .
\label{eq:pseudoinversa-AAeEXTinversivel-2}
\end{equation} 
For instance, for $A=\left(\begin{smallmatrix} 2 & 0 & i \\ 0 & i & 1
  \end{smallmatrix}\right)$ one can check that $AA^*$ is invertible,
but $A^* A$ is not, and we have $A^+ = A^*\big(AA^*\big)^{-1}=\frac{1}{9}
\left(\begin{smallmatrix} 4 & -2i \\ 1 & -5i\\ -i & 4
  \end{smallmatrix}\right)$.
Similarly, for $A=\left(\begin{smallmatrix} 1 & 2 \\ 0 & i \\ 0 &
    3 \end{smallmatrix}\right)$, $AA^*$ is singular, but $A^* A$ is
invertible and we have $A^+ = \big(A^*A\big)^{-1}A^* = \frac{1}{10}
\left(\begin{smallmatrix} 10 & 2i & -6 \\ 0 & -i & 3
  \end{smallmatrix}\right)$.

The relations
(\ref{eq:pseudoinversa-AAeEXTinversivel-1})--(\ref{eq:pseudoinversa-AAeEXTinversivel-2})
are significant because they will provide an important hint to find
the Moore-Penrose pseudoinverse of a general matrix, as we will
discuss later. In Proposition
\ref{prop:pseudoinversaeproddematrizespositivas} we will show that one
has in general $A^+ = A^* \big(AA^*\big)^+ = \big(A^* A\big)^+ A^*$
and in Theorem \ref{teor:regularizacaodeTychonovparaAMAIS} we will
discuss what can be done in the cases when $A^*A$ or $A^*A$ are not
invertible.

\section{Further Properties of the Moore-Penrose Pseudoinverse}
\label{sec:OutrasPropriedadesdaPseudo-InversadeMoore-Penrose}

The following properties of the Moore-Penrose pseudoinverse follow
immediately from its definition and from uniqueness. The proofs are
elementary and left to the reader: for any $A\in\Mat(\C, \; m, \; n)$
one has
\begin{enumerate}
\item $\big(A^+\big)^+=A$,
\item $\big(A^+\big)^T=\big(A^T\big)^+$,
  $\overline{A^+}=\left(\overline{A}\right)^+$ and, consequently
  $\big(A^+\big)^*=\big(A^*\big)^+$,
\item $(z A)^+=z^{-1}A^+$ for all $z\in\C$, $z\neq 0$.
\end{enumerate}
It is however important to remark that for $A\in\Mat(\C, \; m, \; n)$
and $B\in\Mat(\C, \; n, \; p)$, the Moore-Penrose pseudoinverse
$(AB)^+$ is not always equals to $B^+A^+$, in contrast to what happens
with the usual inverse in the case $m=n=p$. A relevant exception will
be found in Proposition
\ref{prop:pseudoinversaeproddematrizespositivas}.

The next proposition lists some important properties that will be
used below.

\begin{proposition}
\label{prop:relacoesuteirpseudoinversaMoorePenrose}
The Moore--Penrose pseudoinverse satisfies the following relations:
\begin{eqnarray}
A^+ & = & A^+ \, \big(A^+\big)^* \, A^* \;, 
\label{eq:ident-pseudoinversa-a}
\\
A & = & A \, A^* \,  \big(A^+\big)^* \; , 
\label{eq:ident-pseudoinversa-b}
\\ 
A^* & = & A^* \, A \, A^+ \; ,
\label{eq:ident-pseudoinversa-c}
\\
A^+ & = & A^* \,  \big(A^+\big)^* \, A^+ \;, 
\label{eq:ident-pseudoinversa-d}
\\  
A & = & \big(A^+\big)^* \, A^* \, A \;,
\label{eq:ident-pseudoinversa-e}
\\ 
A^* & = & A^+ \, A \, A^* \; , 
\label{eq:ident-pseudoinversa-f}
\end{eqnarray}
valid for all $A\in\Mat(\C, \; m, \; n)$.
$\EndofStatement$
\end{proposition}

For us, the most relevant of the relations above is relation 
(\ref{eq:ident-pseudoinversa-c}), since we will make use of it in the
proof of Proposition  \ref{teor:OptimizacaoLinear} we when deal with 
optimization of least squares problems.

\Proof{ of Proposition \ref{prop:relacoesuteirpseudoinversaMoorePenrose}} 
Since $AA^+$ is self-adjoint, one has $AA^+=\big(AA^+\big)^* =
\big(A^+\big)^*A^* $. Multiplying to the left by $A^+$, we get
$A^+=A^+ \big(A^+\big)^* A^*$, proving
(\ref{eq:ident-pseudoinversa-a}).
Replacing $A\to A^+$ and using the fact that $A=\big(A^+\big)^+$, one gets
from (\ref{eq:ident-pseudoinversa-a})  $A=A A^* \big(A^+\big)^*$, which is
relation (\ref{eq:ident-pseudoinversa-b}).
Replacing $A\to A^*$ and using the fact that
$\big(A^*\big)^+=\big(A^+\big)^*$, we get from
(\ref{eq:ident-pseudoinversa-b}) that $A^*=A^* A A^+$, which is
relation (\ref{eq:ident-pseudoinversa-c}).

Relations
(\ref{eq:ident-pseudoinversa-d})--(\ref{eq:ident-pseudoinversa-f}) can
be obtained analogously from the fact that $A^+A$ is also
self-adjoint, but they follow more easily by replacing $A\to A^*$ in
(\ref{eq:ident-pseudoinversa-a})--(\ref{eq:ident-pseudoinversa-c}) and
by taking the adjoint of the resulting expressions.
$\QED$

From Proposition \ref{prop:relacoesuteirpseudoinversaMoorePenrose}
other interesting results  can be obtained, some of which are listed
in the following proposition:

\begin{proposition}
\label{prop:pseudoinversaeproddematrizespositivas}
For all $A\in\Mat(\C, \; m, \; n)$ one has
\begin{equation}
\big(AA^*\big)^+ \; = \; \big(A^*\big)^+ A^+  
\;.
\label{eq:poiynOUIybiyutuyGUg}
\end{equation}
From this we get
\begin{equation}
A^+ \; = \; A^* \big(AA^*\big)^+ \; = \; \big(A^* A\big)^+ A^* \;,
\label{eq:LniuybuyTUYTygfytgHGv}
\end{equation}
also valid for all $A\in\Mat(\C, \; m, \; n)$. 
$\EndofStatement$
\end{proposition}

Expression (\ref{eq:LniuybuyTUYTygfytgHGv}) generalizes
(\ref{eq:pseudoinversa-AAeEXTinversivel-1})--(\ref{eq:pseudoinversa-AAeEXTinversivel-2})
and can be employed to compute $A^+$ provided $\big(AA^*\big)^+ $ or
$\big(A^* A\big)^+$ were previously known.

\Proof{ of Proposition \ref{prop:pseudoinversaeproddematrizespositivas}} 
Let $B=\big(A^*\big)^+ A^+$. One has
$$
AA^* 
\; \stackrel{(\ref{eq:ident-pseudoinversa-b})}{=} \; 
 A \, A^* \,  (A^+)^* \, A^*
\; \stackrel{(\ref{eq:ident-pseudoinversa-f})}{=} \; 
 A \, A^* \,  (A^+)^* \,  A^+ \, A \, A^* 
\; = \; 
(AA^*)B(AA^*) \;,
$$
where we use that  $\big(A^*\big)^+=\big(A^+\big)^*$. One also has
$$
B\; = \; 
\big(A^*\big)^+ A^+
\; \stackrel{(\ref{eq:ident-pseudoinversa-a})}{=} \; 
(A^+)^* \, A^+ \, A  \, A^+
\; \stackrel{(\ref{eq:ident-pseudoinversa-d})}{=} \; 
(A^+)^* \, A^+ \, A \,A^* \,  (A^+)^* \, A^+
\; = \; 
B \big(  A \,A^*\big)B
\;.
$$
Notice that
$$
\big(  A \,A^*\big)B
\; = \; 
\Big( A \,A^* (A^+)^* \Big) A^+ 
\; \stackrel{(\ref{eq:ident-pseudoinversa-c})}{=}
AA^+
$$
which is self-adjoint, by definition. Analogously,
$$
B\big(  A \,A^*\big)
\; = \; 
(A^+)^*  \Big( A^+  A \,A^* \Big)
\; \stackrel{(\ref{eq:ident-pseudoinversa-e})}{=}
(A^*)^+ A^* \; ,
$$
which is also self-adjoint. The facts exposed in the lines above 
prove that  $B$ is the Moore-Penrose pseudoinverse of 
$AA^*$, establishing (\ref{eq:poiynOUIybiyutuyGUg}).
Replacing $A\to A^*$ in (\ref{eq:poiynOUIybiyutuyGUg}), one also gets
\begin{equation}
\big(A^*A\big)^+ \; = \; A^+ \big(A^*\big)^+  \; .
\label{eq:poiynOUIybiyutuyGUg-2}
\end{equation}
Notice now that
$$
A^* \big(AA^*\big)^+
\; \stackrel{(\ref{eq:poiynOUIybiyutuyGUg})}{=}
A^* \big(A^*\big)^+ A^+ 
\; \stackrel{(\ref{eq:ident-pseudoinversa-d})}{=}
A^+
$$
and that
$$
 \big(A^*A\big)^+ A^*
\; \stackrel{(\ref{eq:poiynOUIybiyutuyGUg-2})}{=}
 A^+ \big(A^*\big)^+ A^*
\; \stackrel{(\ref{eq:ident-pseudoinversa-a})}{=}
A^+
\;,
$$
establishing (\ref{eq:LniuybuyTUYTygfytgHGv}).
$\QED$

\topico{The kernel and the range of a matrix and the Moore-Penrose
  pseudoinverse}

The kernel and the range (or image) of a matrix $A\in\Mat(\C, \; m, \;
n)$ are defined by $\Ker(A)\defi \{ u \in \C^n|\; Au=0\}$ and
$\Ran(A)\defi\{Au, \; u \in \C^n\}$, respectively.  It is evident that
$\Ker(A)$ is a linear subspace of $\C^n$ and that $\Ran(A)$ is a
linear subspace of $\C^m$.

The following proposition will be used below, but is interesting by itself.

\begin{proposition}
\label{prop:apseudoinversaeonucleo}
Let $A\in\Mat(\C, \; m, \; n)$ and let us define $P_1 \defi \UM_n - A^+ A
\in\Mat(\C, \; n)$ and $P_2 \defi \UM_m - A A^+ \in\Mat(\C, \; n)$.
Then, the following claims are valid:
\begin{enumerate}
\item \label{item:NucImg-a}
$P_1$ and $P_2$ are orthogonal projectors, that means, they satisfy
  $(P_k)^2=P_k$ and $P_k^* = P_k$, $k=1, \; 2$.
\item \label{item:NucImg-b}
$\Ker(A) = \Ran(P_1)$, $\Ran(A)=\Ker(P_2)$,
$\Ker(A^+)=\Ran(P_2)$ and $\Ran\big(A^+\big)= \Ker(P_1)$.
\item \label{item:NucImg-c}
$\Ran(A)=\Ker\big(A^+\big)^\perp$ and $\Ran\big(A^+\big)=\Ker(A)^\perp$. 
\item \label{item:NucImg-d}
$\Ker(A)\oplus \Ran\big(A^+\big)=\C^n$ and
$\Ker\big(A^+\big)\oplus \Ran(A)=\C^m$, both being direct sums of 
orthogonal subspaces.
$\EndofStatement$
\end{enumerate}
\end{proposition}

\Proof{} Since $AA^+$ and $A^+A$ are self-adjoint, so are $P_1$ and $P_2$.
One also has $(P_1)^2= \UM - 2 A^+ A + A^+ AA^+ A = \UM - 2 A^+ A +
A^+ A=\UM - A^+ A=P_1$ and analogously for $P_2$.  This proved item
\ref{item:NucImg-a}.

Let $x\in\Ker(A)$. Since $\Ran(P_1)$ is a closed linear subspace of of
$\C^n$, the ``Best Approximant Theorem'', Theorem \ref{minimizacao},
and the Orthogonal Decomposition Theorem, Theorem
\ref{teo:TeoremadaDecomposicaoOrtogonal-TluUIgi}, guarantee the
existence of a {\em unique} $z_0\in \Ran(P_1)$ such that
$\|x-z_0\|=\min\big\{ \|x-z\|, \; z\in
\Ran(P_1)\big\}$. Moreover, $x-z_0$ is orthogonal to $\Ran(P_1)$. Hence,
there exists at least one $y_0\in\C^m$ such that $x-P_1 y_0$ is
orthogonal to every element of the form $P_1y$, i.e., $\lan x-P_1
y_0, \; P_1y\ran=0$ for all $y\in\C^m$, what implies $\lan
P_1(x-P_1 y_0), \; y\ran=0$ for all $y\in\C^m$ what, in turn,
implies $P_1(x-P_1 y_0)=0$. This, however, says that $P_1 x
=P_1y_0$. Since $x\in \Ker(A)$, one has $P_1 x = x$ (by the definition
of $P_1$). We therefore proved that if $x\in \Ker(A)$ then $x\in
\Ran(P_1)$, establishing that $\Ker(A)\subset \Ran(P_1)$. On the other
hand, the fact that $AP_1= A\big(\UM - A^+A\big)=A-A=0$ implies
$\Ran(P_1)\subset \Ker(A)$, establishing that $\Ran(P_1)= \Ker(A)$.

If $z\in \Ker(P_1)$, then $z=A^+Az$, proving that $z\in
\Ran\big(A^+\big)$.  This established that $\Ker(P_1)\subset
\Ran\big(A^+\big)$. On the other hand, if $u\in \Ran\big(A^+\big)$
then there exists $v\in\C^m$ such that $u=A^+ v$. Therefore, $P_1u=
\big(\UM_n - A^+ A\big)A^+v=\big(A^+ - A^+AA^+\big)v=0$, proving that
$u\in \Ker(P_1)$ and that $\Ran\big(A^+\big)\subset\Ker(P_1)$. This
established that $\Ker(P_1) = \Ran\big(A^+\big)$.

$P_2$ is obtained from $P_1$ by the substitution $A\to A^+$ (recalling
that $\big(A^+\big)^+ =A$). Hence, the results above imply that $\Ran(P_2) =
\Ker\big(A^+\big)$ and that $\Ker(P_2) = \Ran(A)$.  This proves item
\ref{item:NucImg-b}.

If $M \in\Mat(\C, \; p)$ (with $p\in\N$, arbitrary) is self-adjoint,
that $\lan y, \; Mx\ran=\lan My, \; x\ran$ for all $x, \; y\in
\C^p$. This relation makes evident that
$\Ker(M)=\Ran(M)^\perp$. Therefore, item \ref{item:NucImg-c} follows
from item \ref{item:NucImg-b} by taking $M=P_1$ and $M=P_2$.  Item
\ref{item:NucImg-d} is evident from item \ref{item:NucImg-c}.
$\QED$

\section{Tikhonov's Regularization and Existence Theorem for the
  Moore-Penrose Pseudoinverse}
\label{sec:RegularizacaodeTikhonov}

In (\ref{eq:pseudoinversa-AAeEXTinversivel-1}) and
(\ref{eq:pseudoinversa-AAeEXTinversivel-2}) we saw that if
$\big(AA^*\big)^{-1}$ exists, then $A^+ = A^*\big(AA^*\big)^{-1}$ an
that if $\big(A^*A\big)^{-1}$ exists, then $A^+ =
\big(A^*A\big)^{-1}A^*$.  If those inverses do not exist, there is an
alternative procedure to obtain $A^+$. We know from Proposition
\ref{prop:matrizesinvertiveissaodensasnoespacodetodasasmatrizes} that
even if $\big(AA^*\big)^{-1}$ does not exist, the matrix $AA^*
+\mu\UM$ will be invertible for all non-vanishing $\mu\in\C$ with
$|\mu|$ small enough. Hence, we could conjecture that the expressions
$A^*\big(AA^*+\mu\UM\big)^{-1}$ and $ \big(A^*A+\mu\UM\big)^{-1}A^*$
are well-defined for $\mu\neq 0$ and $|\mu|$ small enough and converge
to $A^+$ when the limit $\mu\to 0$ is taken. As will now show, this
conjecture is correct.

The provisional replacement of the singular matrices $AA^*$ or $A^*A$
by the non-singular ones $AA^*+\mu\UM$ or $A^*A+\mu\UM$ (with $\mu\neq
0$ and $|\mu|$ ``small'') is a regularization procedure known as {\em
  Tikhonov's regularization}. This procedure was introduced by
Tikhonov in \cite{Tikhonov2} (see also \cite{TikhonovArsenin} and,
for historical remarks, \cite{Groetsch-IntegralEquations}) in his
search for uniform approximations for the solutions of Fredholm's
equation of the first kind
\begin{equation}
\int_a^b k(x, \; y)\, u(y)\;dy 
\; = \; 
f(x)
\;,
\label{eq:Fredholm-of-the-first-kind}
\end{equation}
where $-\infty<a<b<\infty$ and where $k$ and $f$ are given functions
satisfying adequate smoothness conditions. In operator form,
(\ref{eq:Fredholm-of-the-first-kind}) becomes $Ku=f$ and $K$ is well
known to be a compact operator (see, e.g., \cite{Reed-Simon-1}) if $k$
is a continuous function. By using the method of finite differences or
by using expansions in terms of orthogonal functions, the inverse
problem (\ref{eq:Fredholm-of-the-first-kind}) can be replaced by an
approximating inverse matrix problem $Ax=y$, like
(\ref{eq:LuybiUYtvtRytfr-00}). By applying $A^*$ to the left, one gets
$A^*Ax=A^*y$. Since the inverse of $A^*A$ may not exist, one first
considers a solution $x_\mu$ of the {\em regularized equation}
$\big(A^*A+\mu\UM\big)x_\mu = A^*y$, with some adequate $\mu\in\C$,
and asks whether the limit $\lim_{|\mu|\to
  0}\big(A^*A+\mu\UM\big)^{-1}A^*y$ can be taken. As we will see, the
limit exists and is given precisely by $A^+y$. In Tikhonov's case, the
regularized equation $\big(A^*A+\mu\UM\big)x_\mu = A^*y$ can be
obtained from a related Fredholm's equation of the \underline{second}
kind, namely $K^*Ku_{\mu} + \mu u_{\mu}=K^*f$, for which the existence
of solutions, i.e., the existence of the inverse $(K^*K+\mu\UM)^{-1}$,
is granted by Fredholm's Alternative Theorem (see, e.g.,
\cite{Reed-Simon-1}) for all $\mu$ in the resolvent set of $K^*K$ and,
therefore, for all $\mu>0$ (since $K^*K$ is a positive compact
operator)\footnote{Tikhonov's argument in \cite{Tikhonov2} is actually
  more complicated, since he does not consider the regularized
  equation $\big(K^*K + \mu\UM \big)u_{\mu}=K^*f$, but a more general
  version where the identity operator $\UM$ is replaced by a
  Sturm-Liouville operator.}. It is then a technical matter to show
that the limit $\displaystyle \lim_{ {\mu\to 0}\atop {\mu>0} }u_{\mu}$
exists and provides a uniform approximation to a solution of
(\ref{eq:Fredholm-of-the-first-kind}).

Tikhonov, however, does not point to the relation of his ideas to the
theory of the Moore-Penrose inverse. This will be described in what
follows.
Our first result, presented in the next two lemmas, establishes that
the limits $\displaystyle \lim_{\mu\to 0} A^*\big(AA^*
+\mu\UM_m\big)^{-1}$ and $\displaystyle \lim_{\mu\to 0} \big(A^*A
+\mu\UM_n\big)^{-1}A^*$, described above, indeed exist and are equal.
\begin{lemma} 
\label{lema:BmueCmusaoiguais}
Let $A\in\Mat(\C, \; m, \; n)$ and let $\mu\in\C$ be such that $AA^*
+\mu\UM_m$ and $A^*A +\mu\UM_n$ are non-singular (that means
$\mu\not\in\sigma\big(AA^*\big)\cup\sigma\big(A^*A\big)$, a finite set). Then,
$A^*\big(AA^* +\mu\UM_m\big)^{-1} = \big(A^*A +\mu\UM_n\big)^{-1}A^*$.
$\EndofStatement$
\end{lemma}

Recall that, by Proposition \ref{prop:OEspectrodeABedeBA},
$\sigma\big(AA^*\big)$ and $\sigma\big(A^*A\big)$ differ at most by
the element $0$.

\Proof{ of Lemma \ref{lema:BmueCmusaoiguais}} 
Let $B_\mu \defi A^*\big(AA^* +\mu\UM_m\big)^{-1}$ and $C_\mu \defi
\big(A^*A +\mu\UM_n\big)^{-1}A^*$.  We have
\begin{multline*}
A^*AB_\mu 
\, = \, 
A^*\big[AA^*\big]\big(AA^* +\mu\UM_m\big)^{-1}
\, = \, 
A^*\big[AA^* +\mu\UM_m-\mu\UM_m\big]\big(AA^* +\mu\UM_m\big)^{-1} 
\\ = \, 
A^*\Big(\UM_m-\mu \big(AA^* +\mu\UM_m\big)^{-1}\Big)
\, = \,
A^*-\mu B_\mu
\, .
\end{multline*}
Therefore,
$\big(A^*A +\mu\UM_n\big)B_\mu=A^*$, what implies
$
B_\mu=\big(A^*A +\mu\UM_n\big)^{-1}A^* =C_\mu
$.
$\QED$

\begin{lemma}
\label{lema:existenciadaregdetichonov}
For all $A\in\Mat(\C, \; m, \; n)$ the limits $\displaystyle
\lim_{\mu\to 0} A^*\big(AA^* +\mu\UM_m\big)^{-1}$ and $\displaystyle
\lim_{\mu\to 0} \big(A^*A +\mu\UM_n\big)^{-1}A^*$ exist and are equal (by 
Lemma \ref{lema:BmueCmusaoiguais}), defining an element of $\Mat(\C,
\; n, \; m)$.
$\EndofStatement$
\end{lemma}

\Proof{} Notice first that $A$ is an identically zero matrix iff
$AA^*$ or $A^*A$ are zero matrices. In fact, if, for instance,
$A^*A=0$, then for any vector $x$ one has $0=\lan x, \;
A^*Ax\ran = \lan Ax, \;Ax\ran=\|Ax\|^2$, proving that
$A=0$. Hence we will assume that $AA^*$ and $A^*A$ are non-zero
matrices.
  
The matrix $AA^*\in\Mat(\C, \; m)$ is evidently self-adjoint. Let
$\alpha_1, \; \ldots , \; \alpha_r$ be its \underline{distinct}
eigenvalues. By the Spectral Theorem for self-adjoint matrices, (see
Theorems \ref{teoremaespectral} and \ref{AutoAdjuntaehDiagonalizavel})
we may write
\begin{equation}
AA^* \; = \; \sum_{a=1}^{r}\alpha_a E_a 
\; ,
\label{eq:wihnpiuYBItryutf}
\end{equation}
where $E_a$ are the spectral projectors of $AA^*$ and satisfy
$E_aE_b=\delta_{ab}E_a$, $E_a^*=E_a$ and $\sum_{a=1}^{r} E_a =\UM_m$. 
Therefore, 
$$
AA^* +\mu\UM_m
\; = \; 
\sum_{a=1}^{r}(\alpha_a + \mu) E_a 
$$
and, hence, for $\mu\not\in\{\alpha_1, \; \ldots , \; \alpha_r\}$, one
has, by (\ref{eq:representacaoespectraldeinvesa}),
\begin{equation}
\big(AA^* +\mu\UM_m\big)^{-1}
\; = \; 
\sum_{a=1}^{r}\frac{1}{\alpha_a + \mu} E_a  
\qquad \mbox{ and }\qquad 
A^*\big(AA^* +\mu\UM_m\big)^{-1}
\; = \; 
\sum_{a=1}^{r}\frac{1}{\alpha_a + \mu} A^*E_a 
\;.
\label{eq:YOIUybyutuygc}
\end{equation}
There are now two cases to be considered: {\em 1.} zero is not an
eigenvalue of $AA^*$ and {\em 2.} zero is eigenvalue of $AA^*$.

In case {\em 1}, it is clear from (\ref{eq:YOIUybyutuygc}) that the
limit $\displaystyle \lim_{\mu\to 0} A^*\big(AA^* +\mu\UM_m\big)^{-1}$
exists and
\begin{equation}
\lim_{\mu\to 0} A^*\big(AA^* +\mu\UM_m\big)^{-1}
\; = \; 
\sum_{a=1}^{r}\frac{1}{\alpha_a} A^*E_a 
\; .
\label{eq:expressaoparaBnocasoAAestnaotemautovalornulo}
\end{equation}

In case {\em 2}, let us have, say, $\alpha_1=0$. The corresponding
spectral projector $E_1$ projects on the kernel of $AA^*$:
$\Ker\big(AA^*\big)\defi \{ u \in \C^n|\; AA^*u=0\}$.  If $x\in
\Ker\big(AA^*\big)$, then $A^* x=0$, because $0=\biglan x, \;
AA^*x\bigran=\lan A^*x, \; A^*x\ran=\big\|A^*x\big\|^2$. Therefore,
\begin{equation}
A^*E_1  \; = \; 0
\label{eq:yfvuyTbihpuybtrdcyr87}
\end{equation}
and, hence, we may write,
$$
A^*\big(AA^* +\mu\UM_m\big)^{-1}
\; = \; 
\sum_{a=2}^{r}\frac{1}{\alpha_a + \mu} A^*E_a 
\; ,
$$
from which we get
\begin{equation}
\lim_{\mu\to 0} A^*\big(AA^* +\mu\UM_m\big)^{-1}
\; = \; 
\sum_{a=2}^{r}\frac{1}{\alpha_a} A^*E_a 
\; .
\label{eq:expressaoparaBnocasoAAesttemautovalornulo}
\end{equation}
This proves that $\displaystyle \lim_{\mu\to 0} A^*\big(AA^*
+\mu\UM_m\big)^{-1}$ always exists.
By Lemma \ref{lema:BmueCmusaoiguais}, the limit $\displaystyle\lim_{\mu\to 0}
\big(A^*A +\mu\UM_n\big)^{-1}A^*$ also exists and coincides with $\displaystyle
\lim_{\mu\to 0} A^*\big(AA^* +\mu\UM_m\big)^{-1}$.
$\QED$

The main consequence is the following theorem, which contains a
general proof for the existence of the Moore-Penrose pseudoinverse:

\begin{theorem}[Tikhonov's Regularization]
\label{teor:regularizacaodeTychonovparaAMAIS}
For all $A\in\Mat(\C, \; m, \; n)$ one has
\begin{equation}
A^+ 
\; = \; 
\lim_{\mu\to 0} A^*\big(AA^* +\mu\UM_m\big)^{-1}
\label{eq:regTichinovparaAMAIS-1}
\end{equation}
and
\begin{equation}
A^+ \; = \; 
\lim_{\mu\to 0} \big(A^*A+\mu\UM_n\big)^{-1}A^*
\; .
\label{eq:regTichinovparaAMAIS-2}
\end{equation}
$\EndofStatement$
\end{theorem}

\Proof{} The statements to be proven are evident if $A=\ZERO_{mn}$
because, as we already saw, $(\ZERO_{mn})^+=\ZERO_{nm}$.  Hence, we
will assume that $A$ is a non-zero matrix. This is equivalent (by the
comments found in the proof o Lemma
\ref{lema:existenciadaregdetichonov}) to assume, that $AA^*$ and
$A^*A$ are non-zero matrices.

By Lemmas \ref{lema:BmueCmusaoiguais} and
\ref{lema:existenciadaregdetichonov} it is enough to prove
(\ref{eq:regTichinovparaAMAIS-1}). There are two cases to be considered:
{\em  1.} zero is not an eigenvalue of
$AA^*$ and {\em 2.} zero is an eigenvalue of $AA^*$.
In case {\em 1.}, we saw in
(\ref{eq:expressaoparaBnocasoAAestnaotemautovalornulo}), 
that
$$
\lim_{\mu\to 0} A^*\big(AA^* +\mu\UM_m\big)^{-1}
\; = \; 
\sum_{a=1}^{r}\frac{1}{\alpha_a} A^*E_a 
\; \ifed \; B 
\;.
$$
Notice now that
\begin{equation}
AB
\; = \; 
\sum_{a=1}^{r}\frac{1}{\alpha_a} AA^*E_a 
\; = \; 
\sum_{a=1}^{r}\frac{1}{\alpha_a} \left( \sum_{b=1}^{r} \alpha_b E_b \right)E_a 
\; = \; 
\sum_{a=1}^{r} \sum_{b=1}^{r} \frac{1}{\alpha_a}\alpha_b\;\delta_{ab}E_a
\; = \; 
\sum_{a=1}^{r} E_a 
\; = \; 
\UM_{m} \;,
\label{eq:KLhniUYiuyYtfygkugfyhygfv-1}
\end{equation}
which is self-adjoint and that
\begin{equation}
BA
\; = \; 
\sum_{a=1}^{r}\frac{1}{\alpha_a} A^*E_a A 
\;,
\label{eq:KLhniUYiuyYtfygkugfyhygfv-2}
\end{equation}
which is also self-adjoint, because $\alpha_a\in\R$ for all $a$ and
because $(A^*E_a A)^* = A^*E_a A$ for all $a$, since $E_a^*=E_a$.

From (\ref{eq:KLhniUYiuyYtfygkugfyhygfv-1}) it follows that
$ABA=A$. From (\ref{eq:KLhniUYiuyYtfygkugfyhygfv-2}) it follows that
$$
BAB
\; = \; 
\left(\sum_{a=1}^{r}\frac{1}{\alpha_a} A^*E_a A\right)
\left(\sum_{b=1}^{r}\frac{1}{\alpha_b} A^*E_b \right)
\; = \; 
\sum_{a=1}^{r}\sum_{b=1}^{r}\frac{1}{\alpha_a\alpha_b} A^*E_a
(AA^*)E_b
\;. 
$$
Now, by the spectral decomposition (\ref{eq:wihnpiuYBItryutf}) for
$AA^*$, it follows that
$(AA^*)E_b=\alpha_b E_b$. Therefore,
$$
BAB
\; = \;
\sum_{a=1}^{r}\sum_{b=1}^{r}\frac{1}{\alpha_a} A^*E_a E_b
\; = \; 
\left( 
\sum_{a=1}^{r}\frac{1}{\alpha_a} A^*E_a
\right)
\bigg(\underbrace{ \sum_{b=1}^{r} E_b}_{\UM_m}\bigg)
\; = \; 
B
\;.
$$
This proves that $A=A^+$ when $0$ is not an eigenvalue of  $AA^*$.

Let is now consider the case when
 $AA^*$ has a zero eigenvalue, say, $\alpha_1$.
As we saw in (\ref{eq:expressaoparaBnocasoAAesttemautovalornulo}), 
$$
\lim_{\mu\to 0} A^*\big(AA^* +\mu\UM_m\big)^{-1}
\; = \; 
\sum_{a=2}^{r}\frac{1}{\alpha_a} A^*E_a 
\; \ifed \; B 
\;.
$$
Using the fact that $(AA^*)E_a=\alpha_a E_a$ (what follows from the
spectral decomposition (\ref{eq:wihnpiuYBItryutf}) for $AA^*$), we get
\begin{equation}
AB
\; = \; 
\sum_{a=2}^{r}\frac{1}{\alpha_a} AA^*E_a 
\; = \; 
\sum_{a=2}^{r}\frac{1}{\alpha_a} \alpha_a E_a 
\; = \; 
\sum_{a=2}^{r} E_a 
\; = \; 
\UM_{m} -E_1 
\;,
\label{eq:KLhniUYiuyYtfygkugfyhygfv-1b}
\end{equation}
which is self-adjoint, since $E_1$ is self-adjoint. We also have
\begin{equation}
BA
\; = \; 
\sum_{a=2}^{r}\frac{1}{\alpha_a} A^*E_a A 
\;,
\label{eq:KLhniUYiuyYtfygkugfyhygfv-2b}
\end{equation}
which is also self-adjoint.

From (\ref{eq:KLhniUYiuyYtfygkugfyhygfv-1b}), it follows that
$ABA=A-E_1 A$.  Notice now that $(E_1 A)^*=A^*E_1=0$, by
(\ref{eq:yfvuyTbihpuybtrdcyr87}). This establishes that $E_1A=0$ and
that $ABA=A$.
From (\ref{eq:KLhniUYiuyYtfygkugfyhygfv-2b}), it follows that
$$
BAB
\; = \; 
\left(\sum_{a=2}^{r}\frac{1}{\alpha_a} A^*E_a A\right)
\left(\sum_{b=2}^{r}\frac{1}{\alpha_b} A^*E_b \right)
\; = \; 
\sum_{a=2}^{r}\sum_{b=2}^{r}\frac{1}{\alpha_a\alpha_b} A^*E_a
(AA^*)E_b
\;. 
$$
Using again $(AA^*)E_b=\alpha_b E_b$, we get
$$
BAB
\; = \;
\sum_{a=2}^{r}\sum_{b=2}^{r}\frac{1}{\alpha_a} A^*E_a E_b
\; = \;
\left( 
\sum_{a=2}^{r}\frac{1}{\alpha_a} A^*E_a
\right)
\underbrace{\left( \sum_{b=2}^{r} E_b\right)}_{\UM_m-E_1}
\; = \; 
B - \sum_{a=2}^{r}\frac{1}{\alpha_a} A^*E_aE_1
\; = \; 
B
\;,
$$
since $E_aE_1=0$ for $a\neq 1$.  This shows that $BAB=B$.  Hence,
we established that $A=A^+$ also in the case when $AA^*$ has a zero
eigenvalue, completing the proof of (\ref{eq:regTichinovparaAMAIS-1}).
$\QED$

\section{The Moore-Penrose Pseudoinverse and 
           the Spectral Theorem}
\label{sec:A-Pseudo-Inversa-de-Moore-Penrose-e-o-Teorema-Espectral}

The proof of Theorem \ref{teor:regularizacaodeTychonovparaAMAIS} also
establishes the following facts:

\begin{theorem}
\label{teor:representacaoquaseespectaldeAMAIS}
Let $A\in\Mat(\C, \; m, \; n)$ be a non-zero matrix and let $AA^*=
\sum_{a=1}^r \alpha_a E_a $ be the spectral representation of $AA^*$,
where $\{\alpha_1 , \; \ldots , \; \alpha_r\}\subset \R$ is the set of
distinct eigenvalues of $AA^*$ and $E_a$ are the corresponding
self-adjoint spectral projections. Then, we have
\begin{equation}
A^+ \; = \; \sum_{{a=1}\atop{\alpha_a\neq 0}}^{r}\frac{1}{\alpha_a} A^*E_a 
\; .
\label{eq:representacaoquaseespectaldeAMAIS-1}
\end{equation}
Analogously, let $A^*A= \sum_{b=1}^s \beta_b F_b $ be the spectral
representation of $A^*A$, where $\{\beta_1 , \; \ldots , \;
\beta_s\}\subset \R$ is the set of distinct eigenvalues of $A^*A$ and
$F_b$ the corresponding self-adjoint spectral projections. Then, we
also have
\begin{equation}
A^+ \; = \; \sum_{{b=1}\atop{\beta_b\neq 0}}^{s}\frac{1}{\beta_b} F_b A^* 
\; .
\label{eq:representacaoquaseespectaldeAMAIS-2}
\end{equation}
Is it worth mentioning that, by Proposition
\ref{prop:OEspectrodeABedeBA}, the sets of non-zero eigenvalues of
$AA^*$ and of $A^*A$ coincide: $\{\alpha_1 , \; \ldots , \;
\alpha_r\}\setminus\{0\} =\{\beta_1 , \; \ldots , \;
\beta_s\}\setminus\{0\}$).

From (\ref{eq:representacaoquaseespectaldeAMAIS-1}) and
(\ref{eq:representacaoquaseespectaldeAMAIS-2}) it follows that for a
non-zero matrix $A$ we have
\begin{eqnarray}
A^+ 
& = &
\sum_{{a=1}\atop{\alpha_a\neq 0}}^{r}
\;\; 
\frac{1}{\alpha_a}
\left(
\prod_{{l=1}\atop {l\neq a}}^r
  \big(\alpha_a-\alpha_l\big)^{-1}
\right)
\;\; A^*
\left[ \prod_{{l=1}\atop {l\neq a}}^r 
\Big( AA^*-\alpha_l\UM_m \Big)\right]
\;,
\label{eq:reppreudoespectraldeAMAIS-1}
\\
A^+ 
& = &
\sum_{{b=1}\atop{\beta_b\neq 0}}^{s}
\;\;
\frac{1}{\beta_b}
\left(
\prod_{{l=1}\atop {l\neq b}}^s
  \big(\beta_b-\beta_l\big)^{-1}
\right)
\;\;
\left[
\prod_{{l=1}\atop {l\neq b}}^s 
\Big( A^*A-\beta_l\UM_n \Big)
\right]
A^*
\; .
\label{eq:reppreudoespectraldeAMAIS-2}
\end{eqnarray}
$\EndofStatement$
\end{theorem}

Expressions (\ref{eq:reppreudoespectraldeAMAIS-1}) or
(\ref{eq:reppreudoespectraldeAMAIS-2}) provide a general algorithm for
the computation of the Moore-Penrose pseudoinverse for any non-zero
matrix $A$. Its implementation requires only the determination of the
eigenvalues of $AA^*$ or of $A^* A$ and the computation of
polynomials on $AA^*$ or $A^* A$.

\Proof{ of Theorem \ref{teor:representacaoquaseespectaldeAMAIS}} Eq.\
 (\ref{eq:representacaoquaseespectaldeAMAIS-1}) was established in
the proof of Theorem
\ref{teor:regularizacaodeTychonovparaAMAIS} (see
(\ref{eq:expressaoparaBnocasoAAestnaotemautovalornulo}) and
(\ref{eq:expressaoparaBnocasoAAesttemautovalornulo})). Relation
(\ref{eq:representacaoquaseespectaldeAMAIS-2}) can be proven
analogously, but it also follows easier (see  
 (\ref{eq:representacaoquaseespectaldeAMAIS-1})), by replacing $A\to
A^*$ and taking the adjoint of the resulting expression.
Relations (\ref{eq:reppreudoespectraldeAMAIS-1}) and
(\ref{eq:reppreudoespectraldeAMAIS-2}) follow from Proposition
\ref{prop:obtendoosprojetoresespectrais-iuybUYyuG}, particularly from
the explicit formula for the spectral projector given in
(\ref{eq:obtendoosprojetoresespectrais-YTvTRVytrvUYtv}).
$\QED$

\section{The Moore-Penrose Pseudoinverse and Least Squares}
\label{sec:A-Pseudo-Inversa-de-Moore-Penrose-e-Problemas-de-Optimizacao-Linear}

Let us now consider one of the main applications of the Moore-Penrose
pseudoinverse, namely, to optimization of linear least squares problems.
Let $A\in\Mat(\C, \; m, \; n)$ and $y\in \C^m$ be given and consider
the problem of finding  $x\in \C^n$ satisfying the linear equation
\begin{equation}
  Ax \; = \; y
\; .
\label{eq:LuybiUYtvtRytfr}
\end{equation}
If $m=n$ and $A$ has an inverse, the (unique) solution is, evidently,
$x=A^{-1}y$.  In the other cases the solution may not exist or may not
be unique.  We can, however, consider the alternative problem of
finding the set of all vectors $x'\in \C^n$ such that the Euclidean
norm $\|Ax'-y\|$ reaches its least possible value.  This set is
called the {\em minimizing set} of the linear problem
(\ref{eq:LuybiUYtvtRytfr}).  Such vectors $x'\in \C^n$ would be the
best approximants for the solution of (\ref{eq:LuybiUYtvtRytfr}) in
terms of the Euclidean norm, i.e., in terms of ``least squares''.  As
we will show, the Moore-Penrose pseudoinverse provides this set of
vectors $x'$ that minimize $\|Ax'-y\|$.  The main result is
condensed in the following theorem:

\begin{theorem}
\label{teor:OptimizacaoLinear}
Let $A\in\Mat(\C, \; m, \; n)$ and $y\in \C^m$ be given. Then, the set
of all vectors of $\C^n$ for which the map $ \C^n \ni x \mapsto
\|Ax-y\|\in[0, \; \infty)$ assumes a minimum coincides with the
set
\begin{equation}
 A^+y + \Ker(A) = \Big\{
  A^+y +\big( \UM_n -A^+A\big) z, \; z\in\C^n \Big\} 
\; .
\label{eq:conjuntominimizante-deoptimizacaolinear}
\end{equation}
By Proposition
\ref{prop:apseudoinversaeonucleo}, we also have $A^+y + \Ker(A)=A^+y +
\Ran\big(A^+\big)^{\perp}$.
$\EndofStatement$
\end{theorem}

Theorem \ref{teor:OptimizacaoLinear} says that the minimizing set of
the linear problem (\ref{eq:LuybiUYtvtRytfr}) consists of all vector
obtained by adding to the vector $A^+y$ an element of the kernel of
$A$, i.e., to all vectors obtained adding to $A^+y$ a vector
annihilated by $A$. Notice that for the elements $x'$ of the
minimizing set of the linear problem (\ref{eq:LuybiUYtvtRytfr}) one
has $\big\|Ax'-y\big\| =
\Big\|\big(AA^+-\UM_m\big)y\Big\| = \|P_2y\|$, which
vanishes if and only if $y\in\Ker(P_2)=\Ran(A)$ (by Proposition
\ref{prop:apseudoinversaeonucleo}), a rather obvious fact.

\Proof{ of Theorem \ref{teor:OptimizacaoLinear}} The image of $A$,
$\Ran(A)$, is a closed linear subspace of $\C^m$.  The Best
Approximant Theorem and the Orthogonal Decomposition Theorem guarantee
the existence of a \underline{unique} $y_0 \in \Ran(A)$ such that
$\|y_0-y\|$ is minimal, and that this $y_0$ is such that
$y_0-y$ is orthogonal to $\Ran(A)$.

Hence, there exists at least one $x_0\in \C^n$ such that
$\|Ax_0-y\|$ is minimal. Such $x_0$ is not necessarily unique
and, as one easily sees, $x_1\in \C^n$ has the same properties if and
only if $x_0 - x_1\in \Ker(A)$ (since $Ax_0=y_0$ and $Ax_1=y_0$, by
the uniqueness of $y_0$).
As we already observed, $Ax_0-y$ is orthogonal to $\Ran(A)$, i.e.,
$\lan (Ax_0 -y), \; Au\ran=0$ for all $u\in \C^n$. This means
that $\Biglan \big(A^*Ax_0-A^*y\big), \; u\Bigran=0$ for all
$u\in \C^n$ and, therefore, $x_0$ satisfies
\begin{equation}
      A^*Ax_0 \; = \;  A^*y\;.    
\label{eq:m5645gouinOIUyuyguy}
\end{equation}
Now, relation (\ref{eq:ident-pseudoinversa-c}) shows us that $x_0=
A^+ y$ satisfies (\ref{eq:m5645gouinOIUyuyguy}), because $A^*AA^+ y
\stackrel{(\ref{eq:ident-pseudoinversa-c})}{=}A^*y$.
Therefore, we conclude that the set of all $x\in \C^n$ satisfying the
condition of $\|Ax-y\|$ being minimal is composed by all
vectors of the form $A^+y + x_1$ with $x_1\in \Ker(A)$.  By
Proposition \ref{prop:apseudoinversaeonucleo}, $x_1$ is of the form
$x_1 = \big(\UM_n -A^+A\big)z$ for some $z\in\C^n$, completing the proof.
$\QED$

%%%%%%%%%%%%%%%%%%%%%%%%%%%%%%%%%%%%%%%%%%%%%%%%%%%%%%%%%%%%%%%%%%%%%%%%%%%%%%%%%
%%%%%%%%%%%%%%%%%%%%%%%%%%%%%%%%%%%%%%%%%%%%%%%%%%%%%%%%%%%%%%%%%%%%%%%%%%%%%%%%%
%%%%%%%%%%%%%%%%%%%%%%%%%%%%%%%%%%%%%%%%%%%%%%%%%%%%%%%%%%%%%%%%%%%%%%%%%%%%%%%%%
%%%%%%%%%%%%%%%%%%%%%%%%%%%%%%%%%%%%%%%%%%%%%%%%%%%%%%%%%%%%%%%%%%%%%%%%%%%%%%%%%
%%%%%%%%%%%%%%%%%%%%%%%%%%%%%%%%%%%%%%%%%%%%%%%%%%%%%%%%%%%%%%%%%%%%%%%%%%%%%%%%%

\begin{appendix}

\begin{center}
\begin{Large}{\bf Appendices}\end{Large}
\end{center}

\section{A Brief Review of Hilbert Space Theory and 
 Linear Algebra}
\label{app:A-Brief-Review-of-Hilbert-Space-Theory-and-Linear-Algebra}

In this appendix we collect the more important definitions and results
on Linear Algebra and Hilbert space theory that we used in the main
part of this paper. For the benefit of the reader, especially of
students, we provide all results with proofs.

\topico{Hilbert spaces. Basic definitions}

A scalar product in a complex vector space $\calV$ is a function
$\calV\times\calV\to\C$, denoted here by $\lan\cdot,\;\cdot\ran$, such
that the following conditions are satisfied: {\em 1.} For all
$u\in\calV$ one has $\lan u, \; u\ran\geq 0$ and $\lan u, \; u\ran=0$
if and only if $u=0$; {\em 2.}  for all $u, \; v_1, \; v_2\in\calV$
and all $\alpha_1, \; \alpha_2\in\C$ one has $\biglan u, \;
(\alpha_1v_1+\alpha_2v_2)\bigran = \alpha_1 \lan u, \;
v_1\ran+\alpha_2 \lan u, \; v_2\ran $ and $\biglan
(\alpha_1v_1+\alpha_2v_2), \; u\bigran = \overline{\alpha_1} \lan v_1,
\; u \ran+\overline{\alpha_2} \lan v_2, \; u \ran $;
{\em 3.}  $\overline{\lan u, \; v\ran}=\lan v, \; u\ran$ for all
$u, \; v\in\calV$.

The norm associated to the scalar product $\lan\cdot,\;\cdot\ran$ is
defined by $\|u\|\defi\sqrt{\lan u,\;u\ran}$, for all $u\in\calV$.  As
one easily verifies using the defining properties of a scalar product,
this norm satisfies the so-called {\em parallelogram identity}: for
all $ a, \; b \in \calV$, one has
\begin{equation}
\| a + b\|^2 + \|a-b\|^2  \; = \; 2 \|a\|^2 + 2 \|b\|^2 \; .
\label{sldkjfbnhv}
\end{equation}

We say that a sequence $\{v_n\in\calV, \;n\in\N \}$ of vectors in
$\calV$ {\em converges} to an element $v\in\calV$ if for all
$\eps>0$ there exists a $N(\eps)\in\N$ such that $\|v_n-v\|\leq \eps$
for all $n\geq N(\eps)$. In this case we write
$v\in\lim_{n\to\infty}v_n$.  A sequence $\{v_n\in\calV, \;n\in\N \}$
of vectors in $\calV$ is said to be a {\em Cauchy sequence} if for all
$\eps>0$ there exists a $N(\eps)\in\N$ such that $\|v_n-v_m\|\leq
\eps$ for all $n, \; m\in\N$ such that $n\geq N(\eps)$ and $m\geq
N(\eps)$.  A complex vector space $\calV$ is said to be a {\em Hilbert
  space} if it has a scalar product and if it is {\em complete}, i.e.,
if all Cauchy sequences in $\calV$ converge to an element of $\calV$.

\topico{The Best Approximant Theorem}

A subset $A$ of a Hilbert space $\calH$ is said to be {\em convex} if
for all $u, \; v\in A$ and all $\mu\in[0, \; 1]$ one has $\mu
u+(1-\mu)v\in A$. A subset $A$ of a Hilbert space $\calH$ is said to
be {\em closed} if every sequence $\{u_n\in A, \; n\in\N \}$ of
elements of $A$ that converges in $\calH$ converges to an element of
$A$. The following theorem is of fundamental importance in the theory of
Hilbert spaces.

\begin{theorem}[Best Approximant Theorem]
\label{minimizacao} 
Let $A$ be a \underline{convex} and \underline{closed} subset of a
Hilbert space $\calH$. Then, for all $ x\in \calH$ there exists a
\underline{unique} $ y\in A$ such that $ \| x - y\|$ equals the
smallest possible distance between $ x$ and $ A$, that means, $ \| x -
y\| = \inf_{y'\in A} \big\|x - y'\big\| $.  $\EndofStatement$
\end{theorem} 

\Proof{} 
Let $ D \geq 0$ be defined by $ D = \inf_{y'\in A} \|x - y'\| $.  For
each $ n\in \N$ let us choose a vector $y_n \in A$ with the property
that
$
\| x - y_n \|^2   <    D^2 + \frac{1}{n}  
$.
Such a choice is always possible, by the definition of the infimum of
a set of real numbers bounded from below.

Let us now prove that the sequence $ y_n$, $n\in\N$ is a Cauchy
sequence in $ \calH$.  Let us take $ a= x - y_n$ and $ b = x-y_m$ in the
parallelogram identity (\ref{sldkjfbnhv}). Then,
$
\big\|2 x - (y_m + y_n)\big\|^2 + \|y_m - y_n\|^2
  = 
2 \|x - y_n\|^2 + 2 \|x-y_m\|^2   
$.
This can be written as 
$
\|y_m - y_n\|^2  = 
2 \|x - y_n\|^2 + 2 \|x-y_m\|^2  -
4 \big\|x - (y_m + y_n)/2\big\|^2  
$.
Now, using the fact that $\| x - y_n \|^2 <  D^2 + \frac{1}{n}$ for each
$n\in\N$, we get
$$
\|y_m - y_n\|^2  \; \leq  \; 
4D^2 + 2\left( \frac{1}{n}+\frac{1}{m}\right)
- 4 \big\|x - (y_m + y_n)/2\big\|^2  \; .
$$
Since $ (y_m + y_n)/2 \in A $ the left hand side is a convex
linear combination of elements of the convex set $A$. Hence, by the
definition of $ D$,
$
\big\|x - (y_m + y_n)/2\big\|^2   \geq   D^2  
$.
Therefore, we have
$$
\|y_m - y_n\|^2  
\; \leq \; 
4D^2 + 2\left( \frac{1}{n}+\frac{1}{m}\right) - 4D^2
\; = \; 
2\left( \frac{1}{n}+\frac{1}{m}\right)
\; .
$$
The right hand side can be made arbitrarily small, by taking both $m$
and $n$ large enough, proving that $\{y_n\}_{n\in \N}$ is a Cauchy
sequence. Since $A$ is a closed subspace of the complete space $
\calH$, the sequence $ \{y_n\}_{n\in \N}$ converges to $y\in A$.

Now we prove that $ \|x-y \|=D$. In fact, for all $n\in\N$ one has
$$
\|x-y \|  
\; = \; 
\big\| (x - y_n)  - (y - y_n) \big\|  
\; \leq \; 
\| x - y_n \|+\|  y - y_n \| 
 \; <  \; 
\sqrt{D^2 + \frac{1}{n}} + \|  y - y_n \|  
\; .
$$
Taking $n\to \infty$ and using the fact that $ y_n$ converges to $ y$,
we conclude that $\|x-y \| \leq D $. One the other hand $\|x-y \| \geq
D $ by the definition of $D$ and we must have $\|x-y \| = D$.

At last, it remains to prove the uniqueness of $y$. Assume that there
is another $y'\in A$ such that $ \big\|x-y' \big\| = D$. Using again the
parallelogram identity (\ref{sldkjfbnhv}), but now with $ a = x-y$ and $
b = x - y'$ we get 
$$
\big\|2 x - (y + y')\big\|^2 + \big\|y - y'\big\|^2
 \; = \; 
2 \big\|x - y\big\|^2 + 2 \big\|x-y'\big\|^2  
 \; = \;  4D^2  \; ,
$$
that means,
$$
\big\|y - y'\big\|^2 \;  = \;  4D^2 - \big\|2 x - (y + y')\big\|^2
 \; = \; 
4D^2 - 4 \Big\| x - \big(y + y'\big)/2 \Big\|^2  \; .
$$
Since $(y + y')/2 \in A $ (for $A$ being convex)
it follows that
$
\big\| x - (y + y')/2\big\|^2    \geq   D^2
$
and, hence, $ \big\|y - y'\big\|^2 \leq 0 $, proving that $y=y'$. 
$ \QED$

\topico{Orthogonal complements}

If $ E$ is a subset of a Hilbert space $ \calH$, we define its
{\em orthogonal complement} $E^\perp$ as the set of of vectors in 
$ \calH$ orthogonal to all vectors in $E$:
$
E^\perp  =  
\Big\{
y \in \calH \big| \;\; \lan y, \; x \ran =0 \mbox{ for all } x\in E
\Big\}
$.
The following proposition is of fundamental importance:
\begin{proposition}
\label{prop:Eperp}
The orthogonal complement $E^\perp$ of a subset $ E$ of a Hilbert space
$ \calH$ is a \underline{closed} linear subspace of $\calH$.
$\EndofStatement$
\end{proposition}

\Proof{} If $ x, \; y \in E^\perp$, then, for any
$\alpha, \; \beta \in \C$, one has 
$
\lan \alpha x + \beta y, \; z \ran   =  
\overline{\alpha} \lan x, \; z\ran + \overline{\beta} \lan y, \; z\ran  =  0 
$
for any $ z\in E$, showing that $\alpha x + \beta y\in
E^\perp $. Hence,  $E^\perp$  is a linear subspace of $\calH$.
If $ x_n$ is a sequence in $E^\perp$ converging to
 $ x \in \calH$, then, for all $ z\in E$ one has
$\displaystyle
\lan x, \; z \ran   =  
\left\langle \lim_{n\to \infty}x_n, \; z \right\rangle   =  
 \lim_{n\to \infty} \lan x_n, \; z \ran   =   0  
%\label{dhsdfYH}
$,
since $\lan x_n, \; z \ran = 0 $ for all $ n$. Hence, $ x\in E^\perp$,
showing that $E^\perp$ is closed.  Above, in the first equality, we
used the continuity of the scalar product.
$\QED$

\topico{The Orthogonal Decomposition Theorem}

\begin{theorem}[Orthogonal Decomposition Theorem]
\label{teo:TeoremadaDecomposicaoOrtogonal-TluUIgi}
  Let $ \calM $ be a closed and linear (and therefore convex) subspace
  of a Hilbert space $ \calH$. Then every $x \in \calH$ can be written
  in a unique way in the form $ x = y+z $, with $ y\in \calM$ and $ z
  \in \calM^\perp$.  The vector $y$ is such that $ \|x -y\| =
  \inf_{y'\in \calM}\big\|x - y'\big\|$, i.e., is the best approximant of $x$
  in $ \calM $.
$\EndofStatement$
\end{theorem}

\Proof{} Let $x$ be an arbitrary element of $\calH$.  Since $\calM$ is
convex and closed, let us evoke Theorem \ref{minimizacao} and 
choose $ y$ as the (unique) element of $ \calM$ such that $ \|x -y\| =
\inf_{y'\in \calM}\big\|x - y'\big\|$. Defining
$z \defi x-y$ all we have to do is to show that $ z\in \calM^\perp$
and to show uniqueness of $y$ and $z$.
Let us first prove that $z\in
\calM^\perp$. By the definition of $y$ one has
$
\| x - y\|^2   \leq  \big\| x - y -\lambda y'\big\|^2
$
for all $\lambda \in \C$ and all $y'\in \calM$. By the definition of
$z$, it follows that 
$
\| z\|^2 \;  \leq  \; \big\| z -\lambda y'\big\|^2
$ 
for all $\lambda \in \C$. Writing the right hand side as
$ \biglan z -\lambda y' , \;  z -\lambda y'\bigran $ we get,
$
\| z\|^2  
\leq  
\| z\|^2 - 2 \mathrm{Re}\big(\lambda \lan z, \; y'\ran \big) 
+ |\lambda|^2 \big\|y'\big\|^2  
$.
Hence,
\begin{equation}
2 \mathrm{Re}\big(\lambda \lan z, \; y'\ran \big)  
\; \leq  \; 
|\lambda|^2 \big\|y'\big\|^2 
\; .
\label{sdrvbwert}
\end{equation}
Now, write $\big\langle z, \; y'\big\rangle = \big|\lan z, \;
y'\ran\big|e^{i\alpha} $, for some $ \alpha\in\R$. Since
(\ref{sdrvbwert}) holds for all $ \lambda \in \C$, we can pick $
\lambda$ in the form $ \lambda = te^{-i\alpha} $, $t > 0$ and
(\ref{sdrvbwert}) becomes $ 2 t \big|\lan z, \; y'\ran\big| \leq t^2
\big\|y'\big\|^2 $.  Hence, $ \big|\lan z, \; y'\ran\big| \leq
\frac{t}{2} \big\|y'\big\|^2 $, for all $t>0$. But this is only
possible if the left hand side vanishes: $\big|\lan z, \; y'\ran\big|
=0 $. Since $ y'$ is an arbitrary element of $ \calM$, this shows that
$ z\in \calM^\perp$.

To prove uniqueness, assume that $ x = y' + z'$ with $ y'\in \calM$
and $ z'\in \calM^\perp$. We would have $y - y'= z'- z$. But $y -
y'\in \calM$ and $z'- z \in \calM^\perp$. Hence, both belong to
$\calM\cap \calM^\perp=\{0\}$, showing that $y - y'= z'- z=0$. 
$\QED$

\topico{The spectrum of a matrix}

The {\em spectrum} of a matrix $ A \in \Mat (\C, \; n)$, denoted by $
\sigma(A)$, is the set of all $ \lambda \in \C$ for which the matrix
$\lambda\UM -A $ has no inverse.

The {\em characteristic polynomial} of a matrix $ A \in \Mat (\C, \;
n)$ is defined by $ p_A(z) \defi \det (z\UM -A) $. It is clearly a
polynomial of degree $n$ on $z$.  It follows readily from these
definitions that $\sigma(A)$ coincides with the roots of $ p_A$.  The
elements of $\sigma(A)$ are said to be the {\em eigenvalues} of $A$.
If $\lambda$ is an eigenvalue of $A$, the matrix $A -\lambda\UM$ has
no inverse and, therefore, there exists at least one non-vanishing
vector $v\in\C^n$ such that $(A -\lambda\UM)v=0$, that means, such
that $Av=\lambda v$. Such a vector is said to be an {\em eigenvector}
of $A$ with eigenvalue $\lambda$. The set of all eigenvectors
associated to a given eigenvalues (plus the null vector) is a linear
subspace of $\C^n$, as one easily sees.

The multiplicity of a root $\lambda$ of the characteristic polynomial
of a matrix $ A \in \Mat (\C, \; n)$ is called the {\em algebraic
  multiplicity} of the eigenvalue $\lambda$. The dimension of the
subspace generated by the eigenvectors associated to the eigenvalues
$\lambda$ is called the {\em geometric multiplicity} of the eigenvalue
$\lambda$.  The {\em algebraic multiplicity} of an eigenvalue is
always larger than or equal to its {\em geometric multiplicity}.

\topico{The neighborhood of singular matrices}

\begin{proposition}
\label{prop:matrizesinvertiveissaodensasnoespacodetodasasmatrizes}
Let $A\in \Mat(\C, \; n)$ be arbitrary and let  $B\in \Mat(\C, \;
n)$ be a non-singular matrix. Then, there exist constants $M_1$ and $M_2$
(depending on $A$ and $B$) with $0 < M_1 \leq M_2$ such that 
$A+\mu B$ is invertible for all $\mu\in \C$ with $0<|\mu|< M_1$ and
for all $\mu\in \C$ with $|\mu|> M_2$.
$\EndofStatement$
\end{proposition}

\Proof{} Since $B$ has an inverse, we may write $A+\mu B=\left(\mu\UM
  + AB^{-1}\right)B$. Hence, $A+\mu B$ has an inverse if and only if
$\mu\UM + AB^{-1}$ is non-singular.

Let $C\equiv - AB^{-1}$ and let $\{\lambda_1 , \; \ldots , \;
\lambda_n\}\subset \C$ be the $n$ not necessarily distinct roots of
the characteristic polynomial $p_C$ of $C$.  If all roots vanish, we
take $M_1=M_2>0$, arbitrary. Otherwise, let us define $M_1\defi\min\{
|\lambda_k|, \; \lambda_k\neq 0\}$ and $M_2\defi\max\{ |\lambda_k|, \;
k=1, \, \ldots , \, n\}$.  Then, the sets $\{\mu\in \C| \;0<|\mu|<
M_1\}$ and $\{\mu\in \C| \;|\mu|> M_2\}$ do not contain roots of $p_C$
and, therefore, for $\mu$ in these sets, the matrix $\mu\UM-C=\mu\UM +
AB^{-1}$ is non-singular.
$\QED$

\topico{Similar matrices}

Two matrices  $A \in \Mat (\C, \; n)$ and $B \in \Mat (\C, \; n)$  are
said to be similar if there is a non-singular matrix 
$P \in \Mat (\C, \; n)$ such that $ P^{-1}AP = B$. 
One has the following elementary fact:

\begin{proposition}
\label{prop:trasfsimpolcarac}
Let $ A$ and $ B\in \Mat(\C, \; n)$ be two similar matrices.  Then
their characteristic polynomials coincide, $p_A=p_B$, and, therefore,
their spectra also coincide, $\sigma(A)=\sigma(B)$, as well as the 
geometric multiplicities of their eigenvalues
$\EndofStatement$
\end{proposition}

\Proof{} Let $P\in\Mat(\C, \; n)$ be such that $P^{-1}AP=B$. Then, $
p_A(z) = \det(z \UM -A) = \det\Big(P^{-1}(z \UM -A)P\Big) = \det\big(z \UM
-P^{-1}AP\big) = \det(z\UM -B) = p_B(z) $, for all $z\in\C$.
$\QED$

\topico{The spectrum of products of matrices}

The next proposition contains a non-evident consequence of 
Propositions  \ref{prop:trasfsimpolcarac} and 
\ref{prop:matrizesinvertiveissaodensasnoespacodetodasasmatrizes}.

\begin{proposition}
\label{prop:polcarABigualapolcarBA}
Let $A, \; B\in \Mat(\C, \; n)$. Then, the characteristic polynomials
of the matrices $AB$ and $BA$ coincide: $p_{AB}=p_{BA}$. Therefore,
their spectra also coincide, $\sigma(AB)=\sigma(BA)$, as well as the
geometric multiplicities of their eigenvalues.
$\EndofStatement$
\end{proposition}

\Proof{} IF $A$ or $B$ (or both) are non-singular, then $AB$ and $BA$
are similar. In fact, in the first case we can write $AB=A(BA)A^{-1}$
and in the second one has $AB=B^{-1}(BA)B$.  In both cases the claim
follows from Proposition \ref{prop:trasfsimpolcarac}. Let us now
consider the case where neither $A$ nor $B$ are invertible.  We know
from Proposition
\ref{prop:matrizesinvertiveissaodensasnoespacodetodasasmatrizes}, that
there exists $M>0$ such that $A+\mu \UM$ is non-singular for all
$\mu\in\C$ with $0<|\mu|<M$. Hence, for such values of $\mu$, we have
by the argument above that $p_{(A+\mu\UM)B}=p_{B(A+\mu\UM)}$. Now the
coefficient of the polynomials $p_{(A+\mu\UM)B}$ and
$p_{B(A+\mu\UM)}$ are polynomials in $\mu$ and, therefore, are
continuous. Hence, the equality $p_{(A+\mu\UM)B}=p_{B(A+\mu\UM)}$
remains valid by taking the limit $\mu\to 0$, leading to
$p_{AB}=p_{BA}$.
$\QED$

Proposition \ref{prop:polcarABigualapolcarBA} can be extended to
products of non-square matrices:

\begin{proposition}
\label{prop:OEspectrodeABedeBA}
Let $A\in \Mat(\C, \; m, \; n)$ and $B\in \Mat(\C, \; n, \;
m)$. Clearly, $AB \in \Mat(\C, \; m)$ and $BA \in \Mat(\C, \;
n)$. Then, one has $x^n p_{AB}(x)=x^m p_{BA}(x)$. Therefore,
$\sigma(AB)\setminus\{0\}=\sigma(BA)\setminus\{0\}$, i.e., the set of
non-zero eigenvalues of $AB$ coincide with the set of non-zero
eigenvalues of $BA$.
$\EndofStatement$
\end{proposition}

\Proof{}
Consider the two $(m+n)\times (m+n)$ matrices defined by
$$
A' 
\; \defi \; 
\begin{pmatrix}
A & \ZERO_{m, \; m}
\\
\ZERO_{n, \; n} & \ZERO_{n, \; m}
\end{pmatrix}
\qquad\mbox{ and }\qquad
B' 
\; \defi \; 
\begin{pmatrix}
B & \ZERO_{n, \; n}
\\
\ZERO_{m, \; m} & \ZERO_{m, \; n}
\end{pmatrix}
\; \; .
$$
See (\ref{eq:MatrizesIJeAlinha-5}).  It is easy to see that
$$
A'B'
\; = \; 
\begin{pmatrix}
AB & \ZERO_{m, \; n}
\\
\ZERO_{n, \; m} & \ZERO_{n, \; n}
\end{pmatrix}
\qquad\mbox{ and that }\qquad
B'A'
\; = \; 
\begin{pmatrix}
BA & \ZERO_{n, \; m}
\\
\ZERO_{m, \; n} & \ZERO_{m, \; m}
\end{pmatrix}
\; .
$$
From this, it is now easy to see that $p_{A'B'}(x)=x^n p_{AB}(x)$ and
that $p_{B'A'}(x)=x^m p_{BA}(x)$. By Proposition
\ref{prop:polcarABigualapolcarBA}, one has $p_{A'B'}(x)=p_{B'A'}(x)$,
completing the proof.
$\QED$

\topico{Diagonalizable matrices}

A matrix $ A \in \Mat (\C, \; n)$ is said to be {\em diagonalizable}
if it is similar to a diagonal matrix. Hence  $ A \in \Mat (\C, \; n)$
is diagonalizable if there exists a non-singular matrix $A \in \Mat
(\C, \; n)$ such that $P^{-1}AP$ is diagonal.
The next theorem gives a necessary and sufficient condition for a
matrix to be diagonalizable:
\begin{theorem}
\label{isuvnijoU}
  A matrix $ A \in \Mat (\C, \; n)$ is diagonalizable if and only if
  it has $n$ linearly independent eigenvectors, i.e., it the subspace
  generated by its eigenvectors is $n$ dimensional.
$\EndofStatement$
\end{theorem}

\Proof{}  
Let us assume that $A$ has $ n$ linearly independent eigenvectors
$ \{v^1, \; \ldots , \; v^n \}$, whose eigenvalues are $\{ d_1 , \;
\ldots , \; d_n \}$, respectively. 
Let 
$ P\in \Mat (\C, \; n)$ be defined by
$
P \; = \; 
\Lcolchete v^1, \; \ldots , \; v^n  \Rcolchete  
$.
By (\ref{BgerV}), one has
$$
A P \; = \; \Lcolchete Av^1, \; \ldots , \; Av^n  \Rcolchete
\; = \;  \Lcolchete d_1 v^1, \; \ldots , \; d_n v^n  \Rcolchete 
$$
and by (\ref{eq:gerVmalD}) one has
$
 \Lcolchete d_1 v^1, \; \ldots , \; d_n v^n  \Rcolchete 
% \; = \; 
% \begin{pmatrix}
% v^1_1 & \cdots & v^n_1 \\
%  \vdots & \ddots & \vdots \\
% v^1_n & \cdots & v^n_n 
% \end{pmatrix} 
% \begin{pmatrix}
% d_1 & \cdots & 0\\
%  \vdots & \ddots & \vdots \\
% 0 & \cdots & d_n
% \end{pmatrix}  
=  
PD 
$.
Therefore $AP = PD$. Since the columns of $P$ are linearly
independent, $P$ is non-singular and one has
$P^{-1}AP = D$, showing that $A$ is diagonalizable.

Let us now assume that $A$ is diagonalizable and that there is a
non-singular $ P\in \Mat (\C, \; n)$ such that
$
P^{-1} A P \; = \; D 
\; = \; 
\diag\big(d_1, \; \ldots , \; d_n\big)
% \begin{pmatrix}
% d_1 & \cdots & 0\\
%  \vdots & \ddots & \vdots \\
% 0 & \cdots & d_n
% \end{pmatrix} 
$.
It is evident that the vectors of the canonical base
(\ref{eq:ABaseCanonica-1}) are eigenvectors of $D$, with $ D{\mathbf
  e}_a = d_a {\mathbf e}_a $.  Therefore, $ v_a = P{\mathbf e}_a$ are
eigenvectors of $ A$, since
$
Av_a  =  A  P{\mathbf e}_a  =  PD {\mathbf e}_a 
 = P\big(d_a{\mathbf e}_a\big)  = 
d_a P{\mathbf e}_a  =  d_a v_a 
$.
To show that these vectors $ v_a$ are linearly independent, assume
that there are complex numbers $\alpha_1, \; \ldots , \; \alpha_n$
such that $ \alpha_1 v_1 + \cdots + \alpha_n v_n = 0 $.  Multiplying
by $P^{-1}$ from the left, we get $ \alpha_1 {\mathbf e}_1 + \cdots +
\alpha_n {\mathbf e}_n = 0 $, implying $\alpha_1 = \cdots =
\alpha_n=0$, since the elements ${\mathbf e}_a$ of the canonical
basis are linearly independent.
$ \Fullbox$

The Spectral Theorem is one of the fundamental results of Functional
Analysis and its version for bounded and unbounded self-adjoint
operators in Hilbert spaces is of fundamental importance for the
so-called probabilistic interpretation of Quantum Mechanics. Here we
prove its simplest version for square matrices.

\begin{theorem}[Spectral Theorem for Matrices] 
\label{teoremaespectral}
A matrix $ A\in \Mat (\C, \; n)$ is diagonalizable if and only if
there exist $ r\in \N$, $1\leq r\leq n$, scalars $ \alpha_1, \; \ldots
, \; \alpha_r\in\C$ and non-zero distinct projectors $E_1 , \; \ldots
, \; E_r \in \Mat (\C, \; n)$ such that
\begin{equation}
       A \; = \; \sum_{a=1}^r \alpha_a E_a  \; ,
\label{decomposicaoespectral}
\end{equation}
and
\begin{equation}
      \UM \; = \; \sum_{a=1}^r E_a 
\;,
\label{OIUHYPoOprimo}
\end{equation}
with
$
               E_i E_j  =  \delta_{i, \, j}E_j  
$.
The numbers $ \alpha_1, \; \ldots , \; \alpha_r$ are the distinct
eigenvalues of $ A$.
$ \EndofStatement$
\end{theorem}

The projectors $ E_a$ in (\ref{decomposicaoespectral}) are called the
{\em spectral projectors} of $ A$. The decomposition
(\ref{decomposicaoespectral}) is called {\em spectral decomposition}
of $ A$.  In Proposition
\ref{prop:obtendoosprojetoresespectrais-iuybUYyuG} we will show how
the spectral projections $ E_a$ can be expressed in terms of
polynomials in $ A$. In Proposition
\ref{prop:unicidadedarepresentacaoespectral} we establish the
uniqueness of the spectral decomposition of a diagonalizable matrix.

\Proof{ of Theorem \ref{teoremaespectral}} If $A\in \Mat (\C, \; n)$
is diagonalizable, there exists $P \in \Mat (\C, \; n)$ such that
$P^{-1}AP = D = \diag (\lambda_1 , \; \ldots , \; \lambda_n)$, where
$\lambda_1 , \; \ldots , \; \lambda_n$ are the eigenvalues of $A$.
Let us denote by $\{ \alpha_1 , \; \ldots , \; \alpha_r\}$, $1 \leq r
\leq n$, the set of all \underline{distinct} eigenvalues of $A$.

One can clearly write
$
D  =  \sum_{a=1}^{r} \alpha_a K_a  
$,
where $K_a\in \Mat (\C, \; n)$ are diagonal matrices having  $0$ or
$1$ as diagonal elements, so that
$$
(K_a)_{ij} \; = \; 
\left\{
\begin{array}{ll}
1 \; , & \mbox{if }\; i=j \;\mbox{ and }\; (D)_{ii} = \alpha_a \; ,\\
0 \; , & \mbox{if }\; i=j \;\mbox{ and }\; (D)_{ii} \neq \alpha_a  \; ,\\
0 \; , & \mbox{if }\; i\neq j \; .
\end{array} 
\right.  
$$
Hence, $(K_a)_{ij}=1$  if $i=j$ and $(D)_{ii} = \alpha_a$ and
$(K_a)_{ij}=0$ otherwise.
It is trivial to see that
\begin{equation}
\sum_{a=1}^{r}  K_a \; = \; \UM
\label{PoiOujjuJ}
\end{equation}
and that
\begin{equation}
K_a K_b \; = \; \delta_{a, \, b}\: K_a .
\label{seifybvOIni}
\end{equation}

Since $A=PDP^{-1}$, one has
$
A  =  \sum_{a=1}^{r} \alpha_a E_a \
$,
where $E_a \defi PK_a P^{-1}$. 
It is easy to prove from (\ref{PoiOujjuJ}) that
$
      \UM  =  \sum_{a=1}^r E_a 
$
and it is easy to prove from (\ref{PoiOujjuJ}) that
$
               E_i E_j  =  \delta_{i, \; j}E_j  
$.

Reciprocally, let us now assume that $A$ has a representation like
(\ref{decomposicaoespectral}), with the $E_a$'s having the above
mentioned properties. Let us first notice that for any vector $x$ and
for $k\in\{1, \; \ldots , \; r\}$, one has by
(\ref{decomposicaoespectral})
$$
AE_k x \; = \; \sum_{j=1}^{r}\alpha_j E_j E_k x  \; = \; 
\alpha_k E_k x   \; .
$$
Hence, $E_k x $ is either zero or is an eigenvalue of $A$. Therefore,
the subspace $\calS$ generated by all vectors $\{E_k x, \; x\in\C^n,
\; k=1, \; \ldots , \; r\}$ is a subspace of the space $\calA$
generated by all eigenvectors of $A$. However, from
(\ref{OIUHYPoOprimo}), one has, for all $x\in\C^n$,
$
x  =  \UM x  =  \sum_{k=1}^{r} E_k x 
$
and this reveals that $\C^n=\calS\subset \calA$. Hence, $ \calA=\C^n$
and by Theorem \ref{isuvnijoU}, $A$ is diagonalizable.
$\Fullbox$

The Spectral Theorem has the following corollary, known as the 
{\em functional calculus}:
\begin{theorem}[Functional Calculus]
\label{teor:CalculoFuncionalparaMatrizes}
  Let $ A \in \Mat (\C, \; n)$ be diagonalizable and let
  $\displaystyle A = \sum_{a=1}^r \alpha_a E_a $ be its spectral
  decomposition. Then, for any polynomial $p$ one has
$\displaystyle
p(A)  =  \sum_{a=1}^r p(\alpha_a) E_a  
$.
$\EndofStatement$
\end{theorem}

\Proof{} By the properties of the spectral projectors $E_a$, one sees
easily that $\displaystyle A^2 = \sum_{a, \; b=1}^r \alpha_a\alpha_b
E_a E_b = \sum_{a, \; b=1}^r \alpha_a\alpha_b \delta_{a, \; b }E_a =
\sum_{a=1}^r (\alpha_a)^2 E_a $.  It is then easy to prove by
induction that $\displaystyle A^m = \sum_{a=1}^r (\alpha_a)^m E_a $,
for all $ m \in \N_0$ (by adopting the convention that $A^0=\UM$, the
case $m=0$ is simply (\ref{OIUHYPoOprimo})). From this, the rest of
the proof is elementary.
$ \Fullbox$ 

One can also easily show that for a non-singular diagonalizable
matrix $A \in \Mat (\C, \; n) $ one has 
\begin{equation}
A^{-1} \; = \; 
\sum_{a=1}^r \frac{1}{\alpha_a} E_a 
\; .
\label{eq:representacaoespectraldeinvesa}
\end{equation}

\topico{Getting the spectral projections}

One of the most useful consequences of the functional calculus is an
explicit formula for the spectral projections of a diagonalizable
matrix $A$ in terms of a polynomial on $A$.

\begin{proposition}
\label{prop:obtendoosprojetoresespectrais-iuybUYyuG}
  Let $A\in \Mat(\C, \; n)$ be non-zero and diagonalizable and let
  $A=\alpha_1 E_1 + \cdots +\alpha_r E_r $ be its spectral
  decomposition. Let the polynomials $p_j$, $j=1, \, \ldots , \, r$,
  be defined by
\begin{equation}
p_j(x) \; \defi \; 
\prod_{{l=1}\atop {l\neq j}}^r 
\left(\frac{x-\alpha_l}{\alpha_j-\alpha_l}\right)
\; .
\label{eq:definicaodospolinoomiospj-ytivuytivTYRyr}
\end{equation}
Then, 
\begin{equation}
E_j 
\; = \; 
p_j(A)
\; = \;
\left( 
       \prod_{{k=1}\atop {k\neq j}}^r \frac{1}{\alpha_j-\alpha_k}
\right)
\prod_{{l=1}\atop {l\neq j}}^r  \Big( A-\alpha_l\UM \Big)
\label{eq:obtendoosprojetoresespectrais-YTvTRVytrvUYtv}
\end{equation}
for all $j=1, \; \ldots , \; r$.
$\EndofStatement$
\end{proposition}

\Proof{} By the definition of the polynomials $p_j$, it is evident
that $p_j(\alpha_k)=\delta_{j, \, k}$. Hence, by Theorem
\ref{teor:CalculoFuncionalparaMatrizes}, $ p_j(A) = \sum_{k=1}^r
p_j(\alpha_k) E_k = E_j $.
$\QED$

\topico{Uniqueness of the spectral decomposition}

\begin{proposition}
  The spectral decomposition of a diagonalizable matrix $A\in
  \Mat(\C, \; n)$ is unique.
  $\EndofStatement$
\label{prop:unicidadedarepresentacaoespectral}
\end{proposition}

\Proof{} Let $\displaystyle A= \sum_{k=1}^r \alpha_k E_k$ be
the spectral decomposition of $A$ as described in Theorem
\ref{teoremaespectral}, where $\alpha_k$, $k=1, \, \ldots ,\, r$, with
$1\leq r\leq n$ are the distinct eigenvalues of $A$, Let
$\displaystyle A= \sum_{k=1}^{s} \beta_k F_k$ be a second
representation of $A$, where the $\beta_k$'s are distinct and where
the $F_k$'s are non-vanishing and satisfy $F_jF_l=\delta_{j, \,
  l}F_l$ and $\displaystyle \UM= \sum_{k=1}^{s} F_k$. For a vector
$x\neq 0$ it holds $x=\sum_{k=1}^{s} F_kx$, so that not all vectors
$F_kx$ vanish. Let $F_{k_0}x\neq 0$.  One has
$AF_{k_0}x=\sum_{k=1}^{s} \beta_k
F_kF_{k_0}x=\beta_{k_0}F_{k_0}x$. This shows that $\beta_{k_0}$
is one of the eigenvalues of $A$ and, hence, $\{ \beta_1, \; \ldots,
\; \beta_{s}\}\subset \{ \alpha_1, \; \ldots, \; \alpha_{r}\}$ and
we must have $s\leq r$. Let us order both sets such that $ \beta_k=
\alpha_k$ for all $1\leq k\leq s$.  Hence,
\begin{equation}
A \; = \; \sum_{k=1}^r \alpha_k E_k \; = \; \sum_{k=1}^{s} \alpha_k F_k
\; .
\label{eq:JboiuyBIuytvIUtyuiyt}
\end{equation}
Now, consider the polynomials $p_j$, $j=1, \, \ldots , \, r$, defined in
(\ref{eq:definicaodospolinoomiospj-ytivuytivTYRyr}), for which
 $p_j(\alpha_j)=1$ and $p_j(\alpha_k)=0$ for all $k\neq j$.
By the functional calculus, it follows from 
(\ref{eq:JboiuyBIuytvIUtyuiyt}) that, for $1\leq j \leq s$,
$$
p_j(A) \; = \; 
\underbrace{\sum_{k=1}^r p_j(\alpha_k) E_k}_{= E_j} 
\; = \; 
\underbrace{\sum_{k=1}^{s} p_j(\alpha_k) F_k}_{= F_j} 
\; ,
\qquad \therefore \quad  E_j\; = \; F_j \; .
$$
(The equality $p_j(A)=\sum_{k=1}^{s} p_j(\alpha_k) F_k$ follows from
the fact that the $E_k$'s and the $F_k$'s satisfy the same algebraic
relations and, hence, the functional calculus also holds for the 
representation of $A$ in terms of the $F_k$'s).  
Since
$\displaystyle \UM = \sum_{k=1}^r E_k = \sum_{k=1}^{s} E_k $, and
$E_j=F_j$ for all $1\leq j \leq s$, one has
$\displaystyle\sum_{k=s+ 1}^r E_k=\ZERO$. Hence, multiplying by
$E_l$, with $s+1\leq l\leq r$, it follows that $E_l = \ZERO$ for all
$s+1\leq l\leq r$. This is only possible if $r=s$, since the $E_k$'s
are non-vanishing. This completes the proof.
$\QED$

\topico{Self-adjointness and diagonalizability}

Let $A \in \Mat (\C, \; m, \; n)$. The adjoint matrix $A^* \in \Mat (\C, \;
n, \; m)$ is defined as the unique matrix for which the equality
$$
\biglan u, \; Av \bigran 
\; = \; 
\biglan A^*u, \; v \bigran 
$$
holds for all $u\in\C^m$ and all $v\in\C^n$. If $A_{ij}$ are the
matrix elements of $A$ in the canonical basis, it is an easy exercise
to show that $\big(A^*\big)_{ij}=\overline{A_{ji}}$, where the bar
denotes complex conjugation. It is trivial to prove that the following
properties hold: {\em 1.} $\big(\alpha_1 A_1 + \alpha_2
A_2\big)^*=\overline{\alpha_1}A_1^*+ \overline{\alpha_2}A_2^*$ for all
$A_1, \; A_2\in \Mat (\C, \; m, \; n)$ and all $\alpha_1, \;
\alpha_2\in\C$; {\em 2.} $\big(AB\big)^*=B^*A^*$ for all $A \in \Mat
(\C, \; m, \; n)$ and $B \in \Mat (\C, \; p, \; m)$; {\em 3.}
$A^{**}\equiv\big(A^*\big)^*=A$ for all $A \in \Mat (\C, \; m, \; n)$.

A square matrix $A \in \Mat (\C, \; n)$ is said to be {\em
  self-adjoint} if $A=A^*$. A square matrix $U \in \Mat (\C, \; n)$ is
said to be {\em unitary} if $U^{-1}=U^*$. Self-adjoint matrices have
real eigenvalues. In fact, if $A$ is self-adjoint,
$\lambda\in\sigma(A)$ and $v\in\C^n$ is a normalized (i.e., $\|v\|=1$)
eigenvector of $A$ with eigenvalue $\lambda$, then
$\lambda=\lambda\lan v, \; v\ran= \lan v, \; \lambda v\ran=\lan v, \;
Av\ran= \lan Av, \; v\ran=\lan \lambda v, \; v\ran=
\overline{\lambda}\lan v, \; v\ran= \overline{\lambda} $, showing that
$\lambda\in\R$.

\topico{Projectors and orthogonal projectors}

A matrix $E\in\Mat(\C, \; n)$ is said to be a {\em projector} if 
 $E^2 =E$ and it is said to be a {\em orthogonal projector} if
it is a self-adjoint projector: $E^2 =E$ and $E^* =E$. 
An important example of an orthogonal projector is the following. 
Let $v\in\C^n$ be such that $\|v\|=1$ and define,
\begin{equation}
   P_v u \; \defi \; \lan v, \; u \ran\: v\; ,
\label{eq:defdePv-IUyboBYtibuyg}
\end{equation}
for each $u\in\C^n$.  In the canonical basis, the matrix elements of
$P_v$ are given by $\big(P_v\big)_{ij}=\overline{v_j}v_i$, where the
$v_k$'s are the components of $v$. One has,
$$
P_v^2 u \; = \; \lan v, \; u \ran\: P_v v 
\; = \; \lan v, \; u \ran\: \lan v, \; v \ran\: v
\; = \; 
\lan v, \; u \ran\: v
\; = \; P_v u \; ,
$$
proving that $P_v^2 = P_v$.  On the other hand, for any 
$a$, $b\in\C^n$ we get
$$
\lan a , \; P_v b \ran \; = \; 
\biglan a, \;  \lan v, \; b \ran\: v \bigran 
\; = \; 
\lan v, \; b \ran\: \lan a, \;  v \ran
\; = \;  
\left\langle\overline{\lan a, \;  v \ran}\: v, \; b \right\rangle
\; = \; 
\biglan \lan v, \;  a \ran\: v, \; b \bigran
\; = \; 
\lan P_v a , \; b \ran \; , 
$$
showing that $P_v^* = P_v$. Another relevant fact is that if
 $v_1$ and $v_2$ are orthogonal unit vectors, i.e.,
 $\lan v_i, \; v_j \ran =\delta_{ij}$, then
$P_{v_1} P_{v_2} = P_{v_2} P_{v_1} =0$.  In fact, for any
$a\in\C^n$ one has
$$
P_{v_1} \big(P_{v_2} a\big) \; = \; P_{v_1} \big(\lan v_2, \; a \ran\: v_2\big) 
\; = \; \lan v_2, \; a \ran\: P_{v_1} v_2 
\; = \; \lan v_2, \; a \ran\: \lan v_1, \; v_2 \ran\: v_1 
\; = \; 
0 \; .
$$ 
This shows that $P_{v_1} P_{v_2}=0$ and, since both are self-adjoint,
one has also $P_{v_2} P_{v_1}=0$.

\topico{Spectral Theorem for self-adjoint matrices}

The following theorem establishes a fundamental fact about
self-adjoint matrices.

\begin{theorem}[Spectral Theorem for Self-adjoint Matrices]
\label{AutoAdjuntaehDiagonalizavel}
If $A \in \Mat (\C, \; n)$ is self-adjoint, one can find a orthonormal
set $\{v_1, \, \ldots , \, v_n\}$ of eigenvectors of $A$ with real
eigenvalues $\lambda_1, \, \ldots , \, \lambda_n$,  respectively, and
one has the spectral representation
\begin{equation}
\label{eq:decomposicaoespectralparaautoadjuntos0}
A  \; = \; \lambda_1 P_{v_1} + \cdots + \lambda_n P_{v_n}  \; ,
\end{equation}
where $P_{v_k}u\defi \lan v_k, \; u\ran v_k$ satisfy
$P_{v_k}^*=P_{v_k}$ and $P_{v_j}P_{v_k}=\delta_{jk}P_{v_k}$ and
one has $\sum_{k=1}^nP_{v_k}=\UM$.

Therefore, if $A \in \Mat (\C, \; n)$ is a self-adjoint matrix it is
diagonalizable. Moreover, there is a \underline{unitary} $P\in\Mat
(\C, \; n)$ such that $P^{-1}AP=\diag\big(\lambda_1, \; \ldots , \;
\lambda_n\big)$.
$\EndofStatement$
\end{theorem}
 
\Proof{}
Let $\lambda_1\in\R$ be an eigenvalue of $A$ and let $v_1$ be a
corresponding eigenvector. Let us choose $\|v_1\|=1$. Define
$A_1\in \Mat (\C, \; n)$ by
$
A_1  \defi   A - \lambda_1 P_{v_1}   
$.
Since both $A$ and $P_{v_1}$ are self-adjoint, so is $A_1$, since
$\lambda_1$ is real.

It is easy to check that $A_1 v_1 =0$. Moreover, $[v_1]^\perp$, the
subspace orthogonal to $v_1$, is invariant under the action of
$A_1$. In fact, for $w \in [v_1]^\perp$ one has $ \lan A_1 w , \; v_1
\ran = \lan w , \;A_1 v_1 \ran = 0 $, showing that $A_1
w\in[v_1]^\perp$.

It is therefore obvious that the restriction of $A_1$ to $[v_1]^\perp$
is also a self-adjoint operator. Let $v_2 \in [v_1]^\perp$ be an
eigenvector of this self-adjoint restriction with eigenvalues
$\lambda_2$ and choose $\|v_2\|=1$. Define
$$
A_2 \; \defi \; A_1 -  \lambda_2 P_{v_2} \; = \;
A - \lambda_1 P_{v_1} - \lambda_2 P_{v_2}  \; .
$$
Since $\lambda_2$ is real, $A_2$ is self-adjoint.
Moreover, $A_2$ annihilates the vectors in the subspace
$[v_1 , \; v_2]$ and keeps $[v_1 , \; v_2]^\perp$ invariant. In fact,
$
A_2 v_1   
=   
A v_1 - \lambda_1 P_{v_1} v_1 - \lambda_2 P_{v_2}v_1  
= 
\lambda_1 v_1 -  \lambda_1 v_1 - \lambda_2 \lan v_2 , \; v_1\ran v_2
 =  0 
$,
since $\lan v_2 , \; v_1\ran =0$. Analogously, 
$
A_2 v_2   =  
 A_1 v_2 - \lambda_2 P_{v_2} v_2  =  
\lambda_2 v_2 -\lambda_2 v_2  = 0 
$.
Finally, for any $\alpha , \; \beta \in \C$ and
$w \in [v_1 , \; v_2]^\perp$ one has
$
\biglan A_2 w , \; (\alpha v_1 + \beta v_2)\bigran  =  
\biglan  w , \; A_2 (\alpha v_1 + \beta v_2)\bigran =  0 
$,
showing that $[v_1 , \; v_2]^\perp$ is invariant by the action of $A_2$.

Proceeding  inductively, we find a set of vectors $\{v_1 ,
\; \ldots , \; v_n\}$, with $\|v_k\|=1$ and with
$ v_a \in [v_1 , \; \ldots , \; v_{a-1}]^\perp
$ for $2\leq a \leq n$, and a set of real numbers
 $\{ \lambda_1 , \; \ldots , \;
\lambda_n \}$ such that
$
A_n  =  A  - \lambda_1 P_{v_1} - \cdots -\lambda_n P_{v_n}
$
annihilates the subspace $ [v_1 , \; \ldots , \; v_{n}]$. But, since
$\{v_1 , \; \ldots , \; v_n\}$ is an orthonormal set, one must have $
[v_1 , \; \ldots , \; v_{n}]=\C^n$ and, therefore, we must have
$A_n=0$, meaning that
\begin{equation}
\label{decomposicaoespectralparaautoadjuntos}
A  \; = \; \lambda_1 P_{v_1} + \cdots + \lambda_n P_{v_n}  \; .
\end{equation}

One has $P_{v_k}P_{v_l} = \delta_{k, \, l}\: P_{v_k}$, since $\lan
v_k, \; v_l\ran=\delta_{kl}$. Moreover, since
$\{v_1 ,
\; \ldots , \; v_n\}$ is a basis in $\C^n$ one has
\begin{equation}
\label{osiPIUYBfO}
x \; = \; \alpha_1 v_1 + \cdots + \alpha_n v_n  
\end{equation}
for all $x\in\C^n$. By taking the scalar product with $v_k$ one gets
that
$\alpha_k=\lan v_k, \; x\ran$ and, hence, 
$$
x \; = \; \lan v_1 , \; x \ran v_1 + \cdots + \lan v_n , \; x \ran v_n
\; = \; 
P_{v_1}x  + \cdots + P_{v_n}x 
\; = \;
\left(
P_{v_1}  + \cdots + P_{v_n}
\right)x \; .
$$
Since $x$ was an arbitrary element of $\C^n$, we established that
$
P_{v_1} + \cdots + P_{v_n} = \UM  
$.

It follows from (\ref{decomposicaoespectralparaautoadjuntos}) that $ Av_a \;
= \; \lambda_a v_a $. Hence, each $v_k$ is an eigenvector of $A$ with
eigenvalue $\lambda_k$. By Theorem \ref{isuvnijoU}, $A$ is
diagonalizable: there is $P\in \Mat (\C, \; n)$ such that 
$P^{-1}AP=\diag\big(\lambda_1, \; \ldots , \; \lambda_n\big)$. As we
saw in the proof of Theorem \ref{isuvnijoU}, we can choose 
$ P = \Lcolchete v^1, \; \ldots , \; v^n \Rcolchete $.
This is, however, a unitary matrix, since, as one easily checks,
$$
P^* P \; = \; 
\begin{pmatrix}
\lan v_1 , \; v_1 \ran & \cdots & \lan v_1 , \; v_n \ran \\
\vdots & \ddots & \vdots \\
\lan v_n , \; v_1 \ran & \cdots & \lan v_n , \; v_n \ran
\end{pmatrix} 
\; = \; \UM
\;,
$$ 
because $\lan v_a , \; v_b \ran = \delta_{a, \, b}$.
$\Fullbox$

\topico{The Polar Decomposition Theorem for square matrices}

It is well-known that every complex number $z$ can be written in the
so-called {\em polar form} $z=|z|e^{i\theta}$, where $|z|\geq 0$ and
$\theta \in [-\pi, \; \pi)$, with $|z|\defi\sqrt{\overline{z}z}$ and
$e^{i\theta} \defi z|z|^{-1}$.  There is an analogous claim for square
matrices $A\in\Mat(\C, \; n)$. This is the content of the so-called
Polar Decomposition Theorem, Theorem
\ref{teo:teoremadadecomposicaopolardematrizes}, below. Let us make
some preliminary remarks.

Let $A\in\Mat(\C, \; n)$ and consider $A^* A$. One has $(A^* A)^* =
A^* A^{**} = A^* A $ and, hence $A^* A$ is self-adjoint. By Theorem
\ref{AutoAdjuntaehDiagonalizavel}, we can find an orthonormal set 
$\{v_k,\; k =1, \, \ldots , \, n\}$ of
eigenvectors of  $A^*
A$, with eigenvalues $d_k, \; k=1, \, \ldots , \, n$, respectively,
with the matrix
\begin{equation}
P\; \defi \; \Lcolchete v_1, \; \ldots , \; v_n \Rcolchete
\label{eq:iuybuytrTYFGJhddd}
\end{equation}
being unitary and such that $P^*\big(A^* A\big)P = D\defi\diag (d_1, \; \ldots
, \;d_n)$. One has $d_k\geq 0$ since $ d_k \|v_k\|^2 =d_k \lan v_k, \;
v_k\ran = \lan v_k, \; Bv_k\ran = \lan v_k, \; A^*Av_k\ran =
\lan Av_k, \; Av_k\ran = \|Av_k\|^2 $ and, hence,
$d_k=\|Av_k\|^2/\|v_k\|^2 \geq 0$.
 
Define $ D^{1/2} \defi \diag \left(\sqrt{d_1}, \; \ldots , \;
  \sqrt{d_n}\right)$. One has $\left(D^{1/2}\right)^2 = D$. Moreover,
$\left(D^{1/2}\right)^*= D^{1/2}$, since every $\sqrt{d_k}$ is real.
The non-negative numbers $\sqrt{d_1}, \; \ldots , \; \sqrt{d_n}$ are
called the {\em singular values} of $A$.

Define the matrix $\sqrt{A^*A}\in\Mat(\C, \; n)$ by
\begin{equation}
\sqrt{A^*A} \; \defi \; PD^{1/2}P^* \; .
\label{eq:definicaoderaizdeAstarA}
\end{equation}
The matrix $\sqrt{A^*A}$ is self-adjoint, since
$\left(\sqrt{A^*A}\right)^*=\left(PD^{1/2}P^* \right)^* =PD^{1/2}P^*=
\sqrt{A^*A}$.  Notice that $\left(\sqrt{A^*A}\right)^2 =
P(D^{1/2})^2P^* =PDP^* =A^*A$.
From this, it follows that
$$
\left(\det\left(\sqrt{A^*A}\right)\right)^2 \; = \;
\det\left(\left(\sqrt{A^*A}\right)^2\right) \; = \; \det(A^*A) \; = \;
\det(A^*)\det(A) \; = \; \overline{\det(A)}\det(A) \; = \;
|\det(A)|^2 \; .
$$
Hence, $\det\left(\sqrt{A^*A}\right)=|\det(A)|$ and, therefore,
$\sqrt{A^*A}$ is invertible if and only if $A$ is invertible.

We will denote $\sqrt{A^*A}$ by $|A|$, following the analogy suggested
by the complex numbers. Now we can formulate the Polar Decomposition
Theorem for matrices:

\begin{theorem}[Polar Decomposition Theorem]
\label{teo:teoremadadecomposicaopolardematrizes}
If $A\in\Mat(\C, \; n)$ there is a matrix $U\in\Mat(\C, \; n)$ such
that
\begin{equation}
A \; = \; U\sqrt{A^*A} \; .
\label{eq:adecomposicaopolardematrizes}
\end{equation}
If $A$ is non-singular, then $U$ is unique. The representation
(\ref{eq:adecomposicaopolardematrizes}) is called the {\em polar
  representation} of $A$.
$\EndofStatement$
\end{theorem}

\Proof{} As above, let $d_k$, $k=1, \, \ldots , \, n$ be the
eigenvalues of $A^*A$ and let $v_k$, $k=1, \, \ldots , \, n$ be a
corresponding orthonormal set of eigenvalues: $A^*Av_k=d_kv_k$ and
$\lan v_k, \; v_l\ran=\delta_{k\, l}$ (see Theorem
\ref{AutoAdjuntaehDiagonalizavel}).

Since $d_k\geq 0$ we order them in a way that $d_k>0$ for all $k=1, \,
\ldots , \, r$ and $d_k=0$ for all $k=r+1, \, \ldots , \, n$.  Hence,
\begin{equation}
Av_k \; = \; 0 \mbox{ for all }  k\; = \; r+1, \; \ldots , \;
n \; ,
\label{eq:sidpfoD3susudygs}
\end{equation}
because $A^*Av_k=0$ implies $0=\lan v_k, \; A^*Av_k\ran=\lan
Av_k, \; Av_k\ran =\|Av_k\|^2$.

For $k=1, \, \ldots , \, r$, let $w_k$ be the vectors defined by
\begin{equation}
w_k \; \defi \; \frac{1}{\sqrt{d_k}}Av_k 
\; , \quad  k\; =\; 1, \, \ldots , \, r
\; .
\label{eq:nkhbfcrtrdtfRR}
\end{equation}
It is easy to see that
$$
\lan w_k, \; w_l\ran \; = \; 
\frac{1}{\sqrt{d_kd_l}}\lan A v_k, \; Av_l\ran
\; = \; 
\frac{1}{\sqrt{d_kd_l}}\lan A^*A v_k, \; v_l\ran
\; = \; 
\frac{d_k}{\sqrt{d_kd_l}}\lan  v_k, \; v_l\ran
\; = \; 
\frac{d_k}{\sqrt{d_kd_l}}\delta_{k\, l}
\; = \; 
\delta_{k\, l} \; ,
$$
for all $k, \; l=1, \, \ldots , \, r$. Hence, 
$\{w_k,\; k =1, \, \ldots , \, r\}$ is an orthonormal set. We can add
to this set an additional orthonormal set $\{w_k,\; k =r+1, \, \ldots
, \, n\}$, in the orthogonal complement of the set generated by 
$\{w_k,\; k =1, \, \ldots , \, r\}$ and get a new 
orthonormal set $\{w_k,\; k =1, \, \ldots , \, n\}$ as a basis for $\C^n$.

Let $P\in \Mat(\C, \; n)$, be defined as in (\ref{eq:iuybuytrTYFGJhddd}) and
let $Q$ and $U$ be elements of $\Mat(\C, \; n)$ defined by
$$
Q \; \defi \; \Lcolchete w_1, \; \ldots , \; w_n \Rcolchete \; ,
\qquad
U  \; \defi \; QP^* \;.
$$
Since $\big\{v_k,\; k =1, \, \ldots , \, n\big\}$ and $\big\{w_k,\; k
=1, \, \ldots , \, n\big\}$ are orthonormal sets, one easily sees that
$P$ and $Q$ are unitary and, therefore, $U$ is also unitary.

It is easy to show that $AP=QD^{1/2}$, where
$D^{1/2}\defi\diag\left(\sqrt{d_1}, \; \ldots , \;
  \sqrt{d_n}\right)$, In fact,
\begin{multline*}
AP \; \stackrel{(\ref{eq:iuybuytrTYFGJhddd})}{=} \; 
A\Lcolchete v_1, \; \ldots , \; v_n \Rcolchete
\; \stackrel{(\ref{BgerV})}{=} \;
\Lcolchete Av_1, \; \ldots , \; Av_n \Rcolchete
\; \stackrel{(\ref{eq:sidpfoD3susudygs})}{=} \;
\Lcolchete Av_1, \; \ldots , \; Av_r\; 0, \; \ldots , \;0 \Rcolchete
\\
\; \stackrel{(\ref{eq:nkhbfcrtrdtfRR})}{=} \;
\Lcolchete \sqrt{d_1}w_1, \; \ldots , \; \sqrt{d_r}w_r\; 0, \; \ldots , \;0 \Rcolchete
\; \stackrel{(\ref{eq:gerVmalD})}{=} \;
\Lcolchete w_1, \; \ldots , \; w_n \Rcolchete D^{1/2} 
\; = \; QD^{1/2}
\; .
\end{multline*}
Now, since $AP=QD^{1/2}$, it follows that $A=QD^{1/2}P^* =
UPD^{1/2}P^* \stackrel{(\ref{eq:definicaoderaizdeAstarA})}{=}U\sqrt{A^*A}$,
as we wanted to show.

To show that $U$ is uniquely determined if $A$ is invertible, assume
that there exists $U'$ such that $A=U\sqrt{A^*A}=U'\sqrt{A^*A}$.  We
noticed above that $\sqrt{A^*A}$ is invertible if and only if $A$ is
invertible. Hence, if $A$ is invertible, the equality
$U\sqrt{A^*A}=U'\sqrt{A^*A}$ implies $U=U'$. If $A$ is not invertible
the arbitrariness of $U$ lies in the choice of the orthonormal set
$\{w_k,\; k =r+1, \, \ldots , \, n\}$. 
$\QED$

The following corollary is elementary:

\begin{theorem}
Let $A\in\Mat(\C, \; n)$. Then, there exists a unitary matrix
$V\in\Mat(\C, \; n)$ such that
\begin{equation}
A \; = \; \sqrt{AA^*}\, V \; .
\label{eq:adecomposicaopolardematrizes-II}
\end{equation}
If $A$ is non-singular, then $V$ is unique.
$\EndofStatement$
\end{theorem}

\Proof{} For the matrix $A^*$, relation
(\ref{eq:adecomposicaopolardematrizes}) says that
$A^*=U_0\sqrt{(A^*)^*A^*}=U_0\sqrt{AA^*}$ for some unitary
$U_0$. Since $\sqrt{AA^*}$ is self-adjoint, one has $A=\sqrt{AA^*}\,
U_0^*$. Identifying $V\equiv U_0^*$, we get what we wanted.
$\QED$

The polar decomposition theorem can be generalized to bounded
or closed unbounded operators acting on Hilbert spaces
and even to $\mathrm{C}^*$-algebras. See e.g., \cite{Reed-Simon-1} and
\cite{BratteliRobinson1}.

\topico{Singular values decomposition}

The Polar Decomposition Theorem, Theorem
\ref{teo:teoremadadecomposicaopolardematrizes}, has a corollary of
particular interest.

\begin{theorem}[Singular Values Decomposition Theorem]
\label{teo:teoremadadecomposicaoemvaloressingularesdematrizes}
Let $A\in\Mat(\C, \; n)$. Then, there exist unitary matrices
$V$ and $W\in\Mat(\C, \; n)$ such that
\begin{equation}
A \; = \; VSW^* \; ,
\label{eq:adecomposicaoemvaloressingularesdematrizes}
\end{equation}
where $S\in\Mat(\C, \; n)$ is a diagonal matrix whose diagonal
elements are the singular values of $A$, i.e., the eigenvalues of
$\sqrt{A^*A}$.  $\EndofStatement$
\end{theorem}

\Proof{} The claim follows immediately from
(\ref{eq:adecomposicaopolardematrizes}) and from
(\ref{eq:definicaoderaizdeAstarA}) by taking $V=UP$, $W=P$ and
$S=D^{1/2}$.  $\QED$

Theorem \ref{teo:teoremadadecomposicaoemvaloressingularesdematrizes}
can be generalized to rectangular matrices. In what follows, $m, \;
n\in\N$ and we will use definitions (\ref{eq:MatrizesIJeAlinha-1}),
(\ref{eq:MatrizesIJeAlinha-5}) and relation
(\ref{eq:MatrizesIJeAlinha-6}), that allows to injectively map
rectangular matrices into certain square matrices.

\begin{theorem}[Singular Values Decomposition Theorem. General Form]
\label{teo:teoremadadecomposicaoemvaloressingularesdematrizes-GERAL}
Let $A\in\Mat(\C, \; m, \; n)$. Then, there exist unitary matrices
$V$ and $W\in\Mat(\C, \; m+n)$ such that
\begin{equation}
A \; = \; I_{m, \; m+n}VSW^*J_{ m+n , \; n} \; ,
\label{eq:adecomposicaoemvaloressingularesdematrizes-GERAL}
\end{equation}
where $S\in\Mat(\C, \; m+n)$ is a diagonal matrix whose diagonal
elements are the singular values of $A'$ (defined in
(\ref{eq:MatrizesIJeAlinha-5})), i.e., are the eigenvalues of
$\sqrt{(A')^*A'}$.  $\EndofStatement$
\end{theorem}

\Proof{} The matrix $A'\in \Mat(\C, \; m+n)$ is a square matrix and,
by Theorem
\ref{teo:teoremadadecomposicaoemvaloressingularesdematrizes}, it can
be written in terms of a singular value decomposition $A'= VSW^*$ with
$V$ and $W\in\Mat(\C, \; m+n)$, both unitary, and $S\in\Mat(\C, \;
m+n)$ being a diagonal matrix whose diagonal elements are the singular
values of $A'$. Therefore,
(\ref{eq:adecomposicaoemvaloressingularesdematrizes-GERAL}) follows
from (\ref{eq:MatrizesIJeAlinha-6}).  
$\QED$

%%%%%%%%%%%%%%%%%%%%%%%%%%%%%%%%%%%%%%%%%%%%%%%%%%%%%%%%%%%%%%%%%%%%%%%%%%%%
%%%%%%%%%%%%%%%%%%%%%%%%%%%%%%%%%%%%%%%%%%%%%%%%%%%%%%%%%%%%%%%%%%%%%%%%%%%%
%%%%%%%%%%%%%%%%%%%%%%%%%%%%%%%%%%%%%%%%%%%%%%%%%%%%%%%%%%%%%%%%%%%%%%%%%%%%

\section{Singular Values Decomposition and 
          Existence of the Moore-Penrose 
          Pseudoinverse}
\label{app:Existencia-e-Decomposicao-em-Valores-Singulares}

We will now present a second proof of the existence of the
Moore-Penrose pseudoinverse of a general matrix $A\in\Mat(\C, \; m
,\; n)$  making use of Theorem
\ref{teo:teoremadadecomposicaoemvaloressingularesdematrizes}.  We
first consider square matrices and later consider general rectangular
matrices.

\topico{The Moore-Penrose pseudoinverse of square matrices}

Let us first consider square diagonal matrices. If $D\in\Mat(\C, \;
n)$ is a  diagonal matrix, its Moore-Penrose pseudoinverse is given by
 $D^+ \in\Mat(\C, \; n)$, where, for $i=1, \; \ldots , \; n$ one has
$$
\big(D^+\big)_{ii} \; = \; 
\left\{
\begin{array}{cl}
\big(D_{ii}\big)^{-1} \;, & \mbox{if } D_{ii}\neq 0\;,
\\
0 \; , & \mbox{if } D_{ii} = 0\;.
\end{array}
\right.
$$
It is elementary to check that $DD^+D=D$, $D^+DD^+=D^+$ and that $DD^+$ and
$D^+D$ are self-adjoint. Actually, $DD^+=D^+D$, a diagonal matrix
whose diagonal elements are either $0$ or $1$:
$$
\big(DD^+\big)_{ii}=\big(D^+D\big)_{ii} \; = \; 
\left\{
\begin{array}{cl}
1 \; , & \mbox{if } D_{ii}\neq 0\;,
\\
0  \; , & \mbox{if } D_{ii} = 0\;.
\end{array}
\right.
$$

Now, let $A\in\Mat(\C, \; n)$ and let $A=VSW^*$ be its singular values
decomposition (Theorem
\ref{teo:teoremadadecomposicaoemvaloressingularesdematrizes}). We
claim that its Moore-Penrose pseudoinverse $A^+$ is given by
\begin{equation}
A^+ \; = \; WS^+V^* \; .
\end{equation}
In fact,
$AA^+A=\big(VSW^*\big)\big(WS^+V^*\big)\big(VSW^*\big)=VSS^+SW^+=VSW^*=A$
and
$$
A^+AA^+ \; = \; \big(WS^+V^*\big)\big(VSW^*\big)\big(WS^+V^*\big) \; =
\; WS^+SS^+V^* \; = \; WS^+V^* \; = \; A^+ \; .
$$ 
Moreover,
$AA^+=\big(VSW^*\big)\big(WS^+V^*\big)=V\big(SS^+\big)V^*$ is
self-adjoint, since $SS^+$ is a diagonal matrix with diagonal elements
$0$ or $1$.  Analogously,
$A^+A=\big(WS^+V^*\big)\big(VSW^*\big)=W\big(S^+S\big)W^*$ is
self-adjoint.

\topico{The Moore-Penrose pseudoinverse of rectangular matrices}

Consider now $A\in \Mat(\C, \; m, \; n)$ and
let $A'\in \Mat(\C, \; m+n)$ be the $(m+n)\times (m+n)$
defined in (\ref{eq:MatrizesIJeAlinha-5}).  Since $A'$ is a square
matrix it has, by the comments above, a unique Moore-Penrose pseudoinverse 
$( A')^+$ satisfying
\begin{enumerate}
\item $A'\big(A'\big)^+A'=A'$,
\item $\big(A'\big)^+A'\big(A'\big)^+=\big(A'\big)^+$,
\item $A'\big(A'\big)^+$ and $\big(A'\big)^+A'$ are self-adjoint.
\end{enumerate}
In what follows we will show that  $A^+\in \Mat(\C, \; n, \; m)$ is
given by
\begin{equation}
A^+ 
\; \defi \;
I_{n,\; m+n } \big(A'\big)^+ J_{m+n, \; m}
\;,
\label{eq:apseudoinversaparamatrizesretangulares}
\end{equation}
with the definitions
(\ref{eq:MatrizesIJeAlinha-1})--(\ref{eq:MatrizesIJeAlinha-2}), i.e.,
\begin{equation}
A^+ 
\; = \;
I_{ n, \; m+n}
\Big( J_{m+n, \; m} A I_{ n, \; m+n } \Big)^+ 
J_{m+n, \; m}
\;.
\label{eq:apseudoinversaparamatrizesretangulares-2}
\end{equation}

The starting point is the existence of the Moore-Penrose
pseudoinverse of the square matrix $A'$. Relation
$A'\big(A'\big)^+A'=A'$ means, using definition 
(\ref{eq:MatrizesIJeAlinha-5}), that
$
J_{ m+n, \; m} A 
\Big[ I_{ n, \; m+n } \big(A'\big)^+ J_{m+n, \; m}\Big] 
A I_{ n, \; m+n } 
 = 
J_{m+n, \; m} A I_{ n, \; m+n }
$
and from (\ref{eq:MatrizesIJeAlinha-3})--(\ref{eq:MatrizesIJeAlinha-4})
it follows, by multiplying to the left by $I_{m, \; m+n}$ and to the
right by $J_{ m+n , \; n}$, that $ A A^+ A = A $, one of the relations
we wanted to prove.

Relation $\big(A'\big)^+A'\big(A'\big)^+=\big(A'\big)^+$ means, using definition
(\ref{eq:MatrizesIJeAlinha-5}), that
$
\big(A'\big)^+ J_{m+n, \; m} A I_{ n, \; m+n } \big(A'\big)^+
= 
\big(A'\big)^+
$.
Multiplying to the left by $I_{ n, \; m+n }$ and to the right by $
J_{ m+n, \; m}$, this establishes that $ A^+ A A^+ = A^+ $.

Since $A'\big(A'\big)^+$ is self-adjoint, it follows from the
definition (\ref{eq:MatrizesIJeAlinha-5}) that $J_{ m+n, \; m} A
I_{ n, \; m+n } \big(A'\big)^+ $ is self-adjoint, i.e.,
$$ 
J_{m+n, \; m} A I_{ n, \; m+n } \big(A'\big)^+ 
\; = \;
\left( A I_{ n, \; m+n } \big(A'\big)^+\right)^* I_{m, \; m+n} 
\; .
$$  
Therefore, multiplying to left by
$I_{m, \; m+n}$ and to the right by $J_{m+n, \; m}$, it follows from
(\ref{eq:MatrizesIJeAlinha-3}) that
$$
A I_{ n, \; m+n } \big(A'\big)^+ J_{m+n, \; m}
\; = \;
I_{m, \; m+n}\Big( A I_{ n, \; m+n } (A')^+\Big)^*
\; =  \;
\left( A I_{ n, \; m+n } \big(A'\big)^+ J_{m+n, \; m} \right)^*
\; ,
$$
proving that $A A^+$ is self-adjoint

Finally, since $\big(A'\big)^+A'$ is self-adjoint, it follows from
definition (\ref{eq:MatrizesIJeAlinha-5}) that $\big(A'\big)^+
J_{m+n, \; m} A I_{ n, \; m+n }$ is
self-adjoint, i.e.,
$$
\big(A'\big)^+ J_{m+n, \; m} A I_{ n, \; m+n }
\; = \;
J_{ m+n , \; n} \left( \big(A'\big)^+ J_{ m+n, \; m}  A \right)^* 
\;.
$$  
Hence, multiplying to the left by $I_{n, \; m+n}$ and to the right
by $J_{ m+n , \; n}$, if follows from (\ref{eq:MatrizesIJeAlinha-4})
that
$$
I_{n, \; m+n}(A')^+ J_{m+n, \; m} A 
\; = \;
 \left( \big(A'\big)^+ J_{m+n, \; m} A \right)^*J_{ m+n , \; n}
\; = \;
 \left( I_{ n, \; m+n} \big(A'\big)^+ J_{m+n, \; m} A \right)^*
\; ,
$$
establishing that $A^+ A$ is self-adjoint.
This proves that $A^+$ given in
(\ref{eq:apseudoinversaparamatrizesretangulares}) is the Moore-Penrose
pseudoinverse of $A$.

\end{appendix}

%%%%%%%%%%%%%%%%%%%%%%%%%%%%%%%%%%%%%%%%%%%%%%%%%%%%%%%%%%%%%%%%%%%%%%%%%%%%%%%%%
%%%%%%%%%%%%%%%%%%%%%%%%%%%%%%%%%%%%%%%%%%%%%%%%%%%%%%%%%%%%%%%%%%%%%%%%%%%%%%%%%
%%%%%%%%%%%%%%%%%%%%%%%%%%%%%%%%%%%%%%%%%%%%%%%%%%%%%%%%%%%%%%%%%%%%%%%%%%%%%%%%%
%%%%%%%%%%%%%%%%%%%%%%%%%%%%%%%%%%%%%%%%%%%%%%%%%%%%%%%%%%%%%%%%%%%%%%%%%%%%%%%%%
%%%%%%%%%%%%%%%%%%%%%%%%%%%%%%%%%%%%%%%%%%%%%%%%%%%%%%%%%%%%%%%%%%%%%%%%%%%%%%%%%

\vspace{0.5cm}

\noindent\textbf{Acknowledgments.} 
We are specially grateful to Prof.\ Nestor Caticha for providing us with
some references on applications of the Moore-Penrose pseudoinverse and
of the singular values decomposition.
This work was supported in part by
the Brazilian Agencies CNPq and FAPESP.\\

\end{document}

%%%%%%%%%%%%%%%%%%%%%%%%%%%%%%%%%%%%%%%%%%%%%%%%%%%%%%%%%%%%%%%%%%%%%%%%%%%%%%%%%%%%%%%%%%%%%%%%%%%%%%
%%%%%%%%%%%%%%%%%%%%%%%%%%%%%%%%%%%%%%%%%%%%%%%%%%%%%%%%%%%%%%%%%%%%%%%%%%%%%%%%%%%%%%%%%%%%%%%%%%%%%%
%%%%%%%%%%%%%%%%%%%%%%%%%%%%%%%%%%%%%%%%%%%%%%%%%%%%%%%%%%%%%%%%%%%%%%%%%%%%%%%%%%%%%%%%%%%%%%%%%%%%%%
%%%%%%%%%%%%%%%%%%%%%%%%%%%%%%%%%%%%%%%%%%%%%%%%%%%%%%%%%%%%%%%%%%%%%%%%%%%%%%%%%%%%%%%%%%%%%%%%%%%%%%
%%%%%%%%%%%%%%%%%%%%%%%%%%%%%%%%%%%%%%%%%%%%%%%%%%%%%%%%%%%%%%%%%%%%%%%%%%%%%%%%%%%%%%%%%%%%%%%%%%%%%%
%%%%%%%%%%%%%%%%%%%%%%%%%%%%%%%%%%%%%%%%%%%%%%%%%%%%%%%%%%%%%%%%%%%%%%%%%%%%%%%%%%%%%%%%%%%%%%%%%%%%%%